\documentclass[aps,prb,superscriptaddress,showpacs,twocolumn]{revtex4}
\usepackage{graphicx}
\usepackage{dcolumn}
\usepackage{bm}
\usepackage{graphicx}
\usepackage{amsmath, amsfonts, amssymb}
\usepackage{amsxtra}
\usepackage{amsthm}

\def\ie{i.e.}

\def\be{\begin{equation}}      
\def\ee{\end{equation}}
\def\beu{\begin{equation*}}   
\def\eeu{\end{equation*}}

\providecommand{\abs}[1]{\left\lvert#1\right\rvert}   
\providecommand{\ket}[1]{\left|#1\right\rangle}

\providecommand{\bra}[1]{\left\langle#1\right|}
\providecommand{\mean}[1]{\left\langle#1\right\rangle}

\providecommand{\del}{\partial}

\providecommand{\M}{M} 
\providecommand{\I}{I} 

\graphicspath{{DNPDDlongfigures/}}

\providecommand{\av}[1]{\left\langle#1\right\rangle}
\renewcommand{\vec}[1]{\mathbf{#1}}
\providecommand{\Bext}{B_\text{ext}}
\newcommand{\s}{d} 
\providecommand{\Tp}{T_+}
\providecommand{\Tm}{T_-}
\providecommand{\TO}{T_0}
\newcommand{\J}{J}
\providecommand{\m}{m_0}
\providecommand{\pf}{p_f}
\providecommand{\Lamw}{\Lambda_+}
\providecommand{\Lams}{\Lambda_0}
\providecommand{\LamR}{\Gamma_R}
\providecommand{\DelO}{\Delta_0}
\providecommand{\Delm}{\Delta_{-}}

\providecommand{\GamO}{\Gamma_0}
\providecommand\Dperp{D_\perp}
\providecommand\Dperpvec{\vec D_\perp}
\providecommand\Dz{D_z}
\providecommand\Dvec{\vec{D}}
\providecommand\Sperp{S_\perp}
\providecommand\Sperpvec{\vec \Sperp}
\providecommand\Sz{S_z}

\providecommand\Svec{\vec{S}}
\providecommand{\zhat}{\hat{z}}
\newcommand{\tred}{\lambda} 
\begin{document}

\title{Preparation of Non-equilibrium Nuclear Spin States in Double Quantum Dots}


\author{M.~Gullans 
}
\affiliation{Department of Physics, Harvard University, Cambridge, MA 02138, USA}
\author{J.~J.~Krich
}
\affiliation{Department of Physics, Harvard University, Cambridge, MA 02138, USA}
\affiliation{Department of Physics, University of Ottawa, Ottawa, ON, Canada}
\author{J.~M.~Taylor}
\affiliation{Joint Quantum Institute, University of Maryland and National Institute of Standards and Technology, College Park, MD} 
\author{B.~I.~Halperin}
\author{M.~D.~Lukin}
\affiliation{Department of Physics, Harvard University, Cambridge, MA 02138, USA}

\date{\today}

\begin{abstract}
We theoretically study  the dynamic polarization of lattice nuclear spins in GaAs double quantum dots containing two electrons.  In our prior work [Phys. Rev. Lett. 104, 226807 (2010)] we identified three regimes of long-term dynamics,  including the build up of a large difference in the Overhauser fields across the dots, the saturation of the nuclear polarization process associated with formation of so-called ``dark states,'' and the elimination of the difference field.  In particular, when the dots are different sizes we found that the Overhauser field becomes larger in the smaller dot. Here we present a detailed theoretical analysis of these problems including a model of the polarization dynamics  and the development of a new numerical method to efficiently simulate semiclassical-central-spin problems.   When nuclear spin noise is included, the results agree with our prior work indicating that large difference fields and dark states are stable configurations, while the elimination of the difference field is unstable; however, in the absence of noise we find all three steady states are achieved depending on parameters.  These results are in good agreement with dynamic nuclear polarization experiments in double quantum dots. 
\end{abstract}
\maketitle

\section{Introduction}

The study of non-equilibrium dynamics of nuclei in solids has a long
history \cite{Abragam78} and has become particularly relevant as
nanoscale engineering and improvements in control allow to probe
mesoscopic collections of nuclear spins
\cite{Yusa05,Dixon97,Salis01,Ono04,Koppens08,Bracker05,Lai06}. This
control has direct applicability to quantum information science, where
nuclear spins are often a main source of dephasing
\cite{Hanson07}. The goal of developing an understanding of
electronic control of nuclei is to circumvent this nuclear dephasing and to turn
nuclear spins into a useful resource  \cite{Klauser08}, as indicated in recent experiments \cite{Reilly08a,Foletti08,Foletti09,Bluhm10,Bluhm11,Shulman12,Frolov12}.

Double quantum dots in III-V semiconductors can be operated with two electrons coupled to approximately $10^4$ to $10^6$ nuclei by the contact hyperfine interaction. Repeated cycles transitioning from the electronic singlet to triplet states can be used to polarize the nuclear spins; electron spin flips between the singlet and triplet spaces occur due to the difference $\Dvec$ in the Overhauser fields on the two dots \cite{Petta08}.  Early experimental \cite{Reilly08a} and theoretical \cite{Ramon07,Ribeiro09,Yao09,Stopa10} work suggested that the polarization process naturally drove the projection of the difference field onto the magnetic field axis $D_z$ to zero.  However, later experiments and theory both showed that the polarization is naturally accompanied by a growth in $D_z$ and that the data in the original experiments showing a suppression in $D_z$ was likely misinterpreted \cite{Foletti09,Gullans10}.  Instead the results are more consistent with the growth of a large $D_z$ accompanied by a reduction in measurement contrast between singlet and triplet states, which makes it appear as if $D_z$ is small \cite{Barthel12}.  

In Ref.\ \onlinecite{Gullans10} we developed a model to describe the long time dynamics of the nuclear spins undergoing adiabatic pumping.   These results are in good agreement with the experiments described above \cite{Foletti09,Shulman12}.  The main conclusion from this work was that when the dots are different sizes the Overhauser field becomes larger in the smaller dot; thereby resulting in large difference fields.  
In the present work, we present a detailed theoretical analysis of these problems.  We  describe the theoretical methods developed to study this system, including a novel  method for efficient simulation of semiclassical central spin problems, and detail the experimentally relevant polarization phenomena we find in our model.   
The main results of the present work are that when nuclear spin noise is included, the more detailed theory presented here agrees with the results of Ref.~\onlinecite{Gullans10}; however, in the absence of nuclear spin noise, states with $D_z=0$ can also be achieved for certain parameters.


Our theoretical methods are based on a semiclassical description of the nuclear spin dynamics in
which the nuclear spins are grouped into small sets, each homogeneously
coupled to the electron spin \cite{Christ07}.  The nuclei in each set may be treated as a single collective spin and a semiclassical treatment is justified provided the number of spins in each set remains large.   Increasing the number of such sets improves the approximation to the true hyperfine coupling.  More formally, we construct  a systematic approximation to the true hyperfine coupling in terms of a reduced set of $M$  coupling constants.    For the optimal choice of coupling constants, we rigorously prove  that  our approximation reproduces the exact semiclassical time dynamics to within a fixed error for a time that increases linearly with $M$.  For large $M$, this allows examination of the  long timescales relevant for polarization experiments.  This approach extends previous work that assumes that all nuclei on a given dot have equal coupling to the electron spin \cite{Ramon07,Ribeiro09,Yao09,Stopa10,Brataas11,Rudner12}; an approach which often incorrectly predicts rapid saturation of the polarization.  Other extensions to this homogenous coupling model, including semiclassical solutions for the central spin \cite{Brataas12,Chen07,Al-Hassanieh06,Loss11}, and cluster and diagramatic expansion
techniques for short time non-equilibrium behavior \cite{Witzel08,Yao06,Coish04} 
do not explore the wide range of time scales or relevant physics for the double dot case.

Our results can be broken up into two distinct cases depending on whether or not the dots are identical.  
When the dots are different sizes, then the hyperfine coupling, which scales inversely with the volume, is larger on the smaller dot and we find that the Overhauser field grows preferentially on the smaller dot as the polarization increases.  This  preferential growth results in a large Overhauser difference field $D_z$.
For two dots with a difference in volume of less than $\sim20\%$ we find a rich and complex phase diagram for the nuclear spin dynamics, which can be broken into two distinct regimes.  The first regime occurs with large external magnetic fields or short cycle times.
In this regime the system saturates without significant polarization because the perpendicular components of $\Dvec$ rapidly approach zero and spin flips are suppressed; the system approaches a semiclassical ``dark state.''
This occurs with no statistical change in the distribution of $\Dz$.  The second regime occurs in the limit of smaller magnetic fields or slower cycle times. 
In this regime, the dynamics are sensitive to the inclusion of nuclear spin noise.  In the absence of nuclear spin noise we find one potential end state of polarization is a ``zero state'' in which all components of $\Dvec \rightarrow 0$.  In this state the singlet and triplet electronic subspaces are completely decoupled and spin flips no longer occur.  Simultaneously, though, there are instabilities leading to the growth of large Overhauser difference fields.  Crucially, when even a small amount of nuclear spin noise is added the zero states strongly destabilize and the system generically becomes unstable to the growth of large difference fields as shown in Ref. \onlinecite{Gullans10}.

%
%
%
%

These results provide a clear picture of the polarization dynamics in such double quantum dot systems and will be a useful guide to future experiments aimed at more precise control of the nuclear spins.  Although the paper is specific to  double quantum dots in GaAs, many of the results and theoretical methods extend to other  central spin systems under investigation \cite{Tarucha11,Steel12,Imamoglu12}.   More generally, this work is of fundamental interest as we explore the dynamics of an interacting, many-body system when it is far from equilibrium \cite{Urbaszek12}.

The paper is organized as follows.  In section II we define the Hamiltonian for the double dot system and introduce the polarization cycle.  In section III we systematically derive a semiclassical model for the nuclear spins starting from the coarse-grained evolution of the nuclear spin density matrix.  In section IV we present our results for identical and unequal dots in the presence  and absence of nuclear spin noise. In appendix A we provide a summary of the parameters used in our simulations.  In appendix B we describe our approach to coarse graining the electron wave function and provide rigorous bounds on the error in time evolution due to the coarse graining.  In appendix C we extend our simulations to the case of multiple nuclear species and find qualitatively the same results as for a single species.


\section{Setup}
\begin{figure}
\includegraphics[width=3.375in]{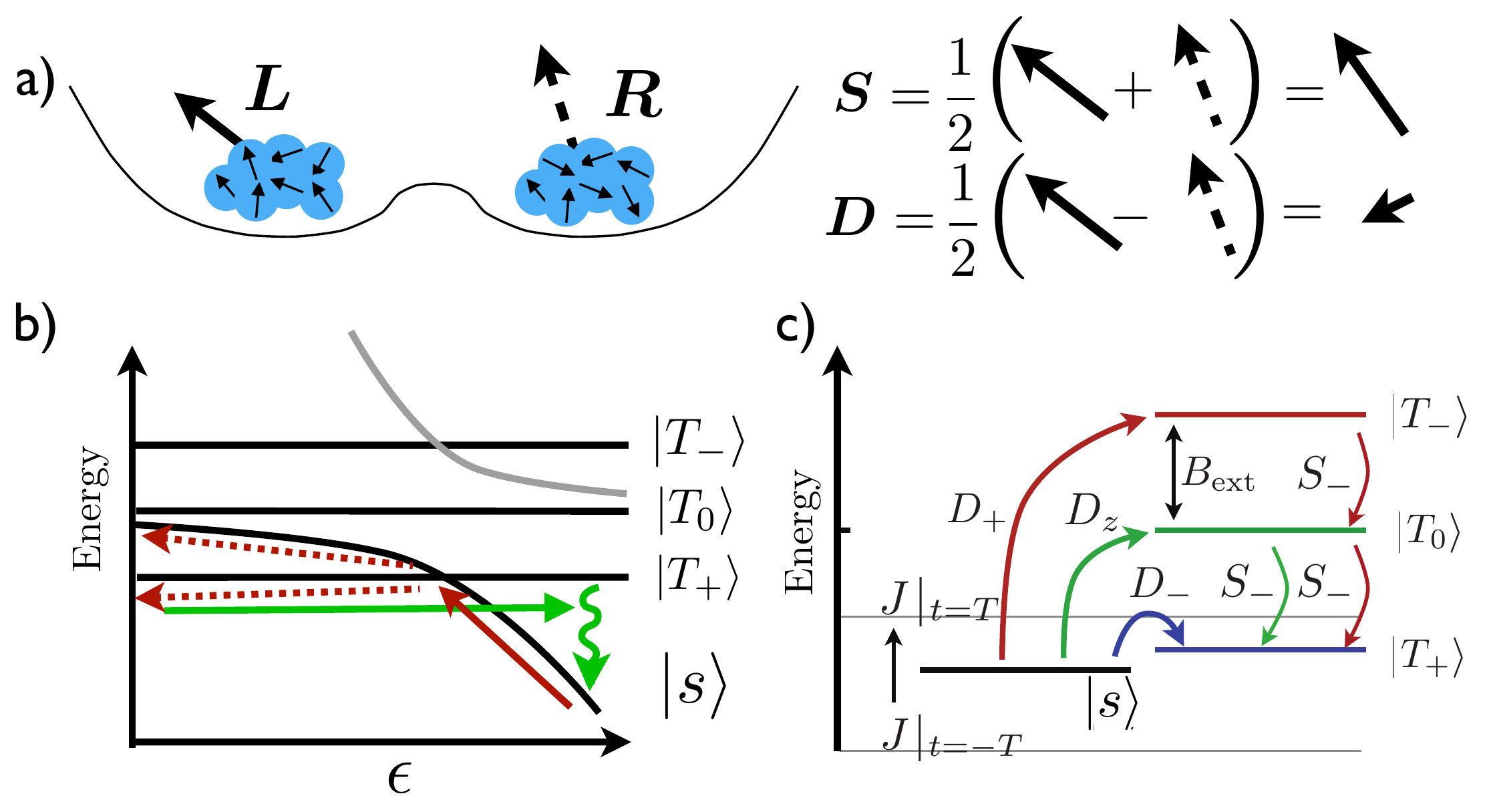}
\caption{\label{fig:sweep} a) The Overhauser field in each dot gives rise to sum and difference fields which are relevant for the double dot system. b) Schematic of two-electron energy levels as a function of detuning $\varepsilon$ between (1,1) and (0,2) charge states. Arrows indicate adiabatic sweep through avoided crossing (pink) and rapid sweep back to (0,2) with reload (green). c) Spin-flip pathways between the $s$ and $\Tp$ states as the exchange energy $J(\varepsilon)$ is swept through the crossing, showing the nuclear operators involved in each path. Each pathway is a term in $\tilde{D}_-$ in Eq.~\ref{eq:Dmtilde}.} 
\end{figure}

For a double quantum dot with two electrons, we can write the Hamiltonian for the lowest energy $(1,1)$ and $(0,2)$ electron states, where $(n,m)$ indicates $n$ ($m$) electrons in the left (right) dot.  To model nuclear polarization, we first derive an effective two-level Hamiltonian to describe the system near the crossing of the singlet $s$ and lowest energy triplet state, $\Tp$, of this two-electron system, then solve the time dynamics.  Dynamic nuclear polarization (DNP) experiments operate near this crossing, typically with an adiabatic sweep of the difference in the dots electric potential through the $s$-$\Tp$ degeneracy (Fig.~\ref{fig:sweep}a), followed by a non-adiabatic return to (0,2) and reset of the electronic state via coupling to leads.

If $\psi_\s(\vec r)$ is the single-particle envelope wave function on dot $\s=l$,$r$ (for the left, right dot), the effective hyperfine coupling for the nuclear spin at $\vec{r}_{k\s}$ is $g_{k\s}=a_{hf}v_0|\psi_\s(\vec r_{k\s})|^2$ where $a_{hf}$ is the hyperfine coupling constant, and $v_0$ is the volume per nuclear spin.  We introduce two collective nuclear spin operators to denote the Overhauser fields in the left ($\hat{\bm{L}}$) and right ($\hat{\bm{R}}$) dots, $\hat{\bm{L}}=\sum_k g_{k l}\vec{I}_{kl}$ and $\hat{\bm{R}}=\sum_k g_{k r}\vec{I}_{k r}$, and further define $\hat{\bm S}=(\hat{\bm L}+\hat{\bm R})/2$, $\hat{\bm D}=(\hat{\bm L}-\hat{\bm R})/2$,
where $\vec{I}_{k\s}$ is the angular momentum of the $k^\text{th}$ nucleus on dot $\s$.  The rms Overhauser energy in the infinite 
temperature ensemble is $\Omega_{d}=(\sum_k g_{kd}^2 I(I+1)/
3)^{1/2}$ where $I$ is the magnitude of each nuclear spin. We define $\Omega = \sqrt{(\Omega_\ell^2+\Omega_r^2)/2}$, and work in energy and magnetic field units such that $\Omega=-\frac{g^*\mu_B}{\hbar}=1$, where $g^*$ is the electron effective g-factor and $\mu_B$ is the Bohr magneton.  
In the basis $\{\ket{s},\ket{\Tp},\ket{\TO},\ket{\Tm}\}$, where the $T_m$ are the $(1,1)$ triplet states and $s$ is the $(1,1)$-$(0,2)$ hybridized singlet state, the Hamiltonian is  \cite{Taylor07} 
\begin{align*}
\begin{split}
  H= \begin{pmatrix}
  -\J(\varepsilon)  & v \hat{D}_+ & -\sqrt{2}  v  \hat{D}_z&- v \hat{D}_- \\
   v\, \hat{D}_-& - \Bext+\hat{S}_z &   \hat{S}_- /\sqrt{2} & 0 \\
  -\sqrt{2} v \hat{D}_z &  \hat{S}_+/\sqrt{2} & 0 & \hat{S}_- /\sqrt{2} \\
  -v \hat{D}_+& 0 &   \hat{S}_+ /\sqrt{2}& \Bext-\hat{S}_z
\end{pmatrix}.
\end{split}
\end{align*}
where $D_\pm \equiv D_x \pm i D_y$ and similarly for $S_\pm$, $\Bext$ is an external magnetic field,  $v=v(\varepsilon)=  \cos \theta(\varepsilon) / \sqrt{2}$, and $\cos \theta(\varepsilon)$ is the overlap of the (1,1) singlet state with the (1,1)-(0,2) hybridized singlet state $\ket{s}$.  The parameters $\cos\theta(\varepsilon)$ and $J(\varepsilon)$, the splitting between $s$ and $T_0$, are both functions of the energy difference $\varepsilon$ between the $(1,1)$ and $(0,2)$ charge states. Here the nuclear spin variables refer to the full quantum mechanical operators on the nuclear spin space.  In appendix C we will consider the case of multiple nuclear species, but for now we consider the nuclei to be spin-$3/2$ of a single species, in a frame rotating at the nuclear Larmor frequency.   

Assuming that $\J,\Bext\gg\Omega$, we perform a formal expansion in the
inverse electron Zeeman energy operator $\hat{m} =\Omega /(\Bext - \hat{S}_z + i
\eta)$ where $\eta>0$ is infinitesimal. We apply a unitary transformation that rotates the
quantization axis of the triplet states to align with $\vec \Bext-\hat{\vec
S}$ and find the Hamiltonian for the $\{ \ket{s},\ket{\Tp} \}$ subspace to first order in $J^{-1}$, $\hat{m}$:
\be
\begin{split}\label{eq:H_STp}
  H_\textrm{eff}=
  \begin{pmatrix}
      -\J(\varepsilon) +\hat{h}_s&  v(\varepsilon) \tilde{D}_+ \\
       v(\varepsilon)\tilde{D}_- &- B_\textrm{ext}+ \hat{h}_T
  \end{pmatrix},
\end{split}
\ee
where the effect of coupling to the higher energy states $\ket{T_0}$ and $\ket{T_-}$ enters as
\begin{align}\label{eq:Dmtilde}
  \hat{h}_s&= -\frac{2 v^2}{\J}\tilde{D}^\dagger_z \tilde{D}_z - \tilde{D}_- \frac{v^2}{\J+\Bext- \hat{S}_z}\tilde{D}_+,\\\nonumber
  \hat{h}_T &= \hat{S}_z  - \frac{1}{4} (\hat{S}_- \hat{S}_+ \hat{m} + \hat{m} \hat{S}_- \hat{S}_+ ),\\\nonumber
  \tilde{D}_- & = \hat{D}_-  +  \hat{m} \hat{S}_- \hat{D}_z - \frac{1}{4} \hat{m}\hat{S}_-^2 \hat{m}\hat{D}_+ - \frac{1}{4} \hat{m}\,\hat{S}_{-} \hat{S}_+\hat{m}\hat{D}_-,\\\nonumber
  \tilde{D}_z &= \hat{D}_z - \frac{1}{2} \big( \hat{S}_+\hat{m}\, \hat{D}_- +  \hat{S}_- \hat{m}\hat{D}_+ \big).
\end{align}
Of particular interest is that the off-diagonal term, which produces nuclear polarization, vanishes in the semiclassical limit of
$\langle \hat{\vec{D}}\rangle \rightarrow 0$, i.e., in the zero states.

\section{Model}
We develop a model for the evolution of the nuclear spin density matrix after one pair of electrons has cycled through the system. We approximate the sweep through the $\ket{s}$-$\ket{\Tp}$ degeneracy as a Landau-Zener process, which we solve approximately for the effect on the nuclear system. By coarse-graining this evolution over a cycle we derive a master equation for the nuclear spins.  Finally, we add the effects of nuclear dipole-dipole interactions and quadrupole splittings phenomenologically.  The derivation presented here is complementary to that of Ref. \onlinecite{Gullans10} and results in the same equations of motion.

The electron system is prepared in $\ket{s}$ at large negative $t=-T/2$, where $T$ is the total cycle time. We identify the (nuclear spin) eigenstates of the operator $\tilde{D}_+ \tilde{D}_-$, labeled $\ket{\Dperp}$ with eigenvalues $D_{\perp}^2$. Since the components of $h_s$ and $h_T$ that do not commute with $\tilde{D}_+ \tilde{D}_-$ are perturbatively small in $m_0$ and $1/J$, we approximate them by keeping only the diagonal components in the two-level-system subspace, sending $ h_{s} \rightarrow \bra{D_{\perp}}  \hat{h}_s \ket{\Dperp}$ and $ h_{T} \rightarrow \bra{D_{\perp}'} \hat h_T \ket{\Dperp'}$  where
  $\smash{\ket{\Dperp'} \equiv D_\perp^{-1}\hat{\tilde{D}}_-\ket{\Dperp}}$.
In this limit, the off-diagonal part of $H_\text{eff}$ in Eq.~\ref{eq:H_STp} produces standard Landau-Zener behavior, while the diagonal components of $H_\text{eff}$ are simply phases picked up by the nuclei, depending on which electronic state is occupied. 
For initial state $\ket{\Psi_0}=\ket{s}\otimes\ket{\Dperp}$, the crossing either leaves the electronic state unchanged
or flips an electron and nuclear spin to the state $\ket{\Tp}\otimes\ket{\Dperp'}$.
We note that $\ket{\Dperp'}$ is an eigenstate of $\tilde{D}_- \tilde{D}_+$ with eigenvalue $D_{\perp}^2$. The problem is now reduced to finding Landau-Zener solutions for each independent two-level system $\{\ket{s}\otimes\ket{\Dperp}$, $\ket{\Tp}\otimes\ket{\Dperp'}\}$.
We model the actual sweep of $\varepsilon$ by a linear sweep of $J$ so
$J(t)=-2\beta^2 t + \Bext$, where $\beta = \sqrt{\frac{1}{2}\abs{d
    \J(\varepsilon)/ d t}\lvert_{t=0}}$. We take $v(\varepsilon)$ to
be constant, valid in the limit of large tunnel coupling, and assume
$\beta\ll\Bext$ to ensure the applicability of
Eq.~\ref{eq:H_STp}.  For moderate magnetic fields $v(\varepsilon) \sim 1/\sqrt{2}$, but it decreases at large magnetic fields as the (1,1)-(0,2) hybridized singlet state has a smaller overlap with (1,1) at the $s$-$T_+$ crossing. 

After one cycle, $\ket{\Psi_0}$ evolves into $\ket{\Psi_1} = c_S \ket{s}\otimes\ket{\Dperp} + c_{T} \ket{\Tp}\otimes\ket{\Dperp'}$.  For $\beta^2 T \gg 1$, the standard Landau-Zener formula gives the flip probability as $p_f = 1 - \exp(-2 \pi \omega^2)$, where $\omega=v \langle \tilde{D}_{\perp} \rangle/\beta$, and
\begin{alignat}{10}
  c_S & =  \sqrt{1 - p_f}  \exp(-i \phi_S), \quad
  c_T  =  \sqrt{p_f}  \exp(-i \phi_T) \nonumber\\\label{eq:phiS}
  \phi_S & \approx  \int_{-T/2}^{T/2} h_{S}  dt \\\nonumber
  \phi_T & \approx  \int_{-T/2}^{t_0} h_{S} dt + (T/2-t_0)
  h_{T}+\phi_{AD}(\omega)\nonumber, 
\end{alignat}
where the crossing occurs at a time $t_0 \approx S_z/\beta^2$. We
include in $\phi_T$ the phase picked up by following the adiabat,
$\phi_{AD}$.  We approximate $\phi_{AD}$ by interpolating between the limits $\omega=v \langle \tilde{D}_\perp \rangle /\beta\rightarrow0$ and $\omega\rightarrow\infty$, giving \cite{Vitanov96}
\begin{align*}
\phi_{AD}=2\pi\omega ^2 + \pf\left\lbrace\omega ^2\left[1-2\pi+\log\left(\frac{\tau^2}{\omega^2}\right)\right]-\pi /4\right\rbrace,
\end{align*}
where $\tau=T\beta/2$.  More accurate approximations can easily be taken into account within our formalism; however we find such corrections have a negligible effect on the long term polarization dynamics because the polarization process rapidly drives $\omega$ to small values.  

We move from the independent two-level systems to the general case by noting that the components of $\ket\Psi$ depend only on the eigenvalue $D_{\perp}$ and on the polarization $S_z$ (which we approximate as commuting). Since the eigenstates of $\tilde{D}_+ \tilde{D}_-$ form a complete basis for the nuclear spin states we can define the complete operator $\hat{p}_f = \sum_{D_{\perp}} p_f(D_{\perp}) \ket{\Dperp} \bra{D_{\perp}}$, and similarly for $\hat \phi_S, \hat \phi_T$. The nuclear spin density matrix after each cycle is given by tracing over the electronic states. The nuclear density matrix evolution is then
\begin{align*}
 \begin{split}
 \rho_{n}&= \sqrt{1-\hat{p}_f} e^{-i \hat{\phi}_S }\rho_{n-1}e^{i \hat{\phi}_S}\sqrt{1-\hat{p}_f}\\
 &+\bigg(\tilde{D}_- \sqrt{\frac{\hat{p}_f}{\tilde{D}_+ \tilde{D}_-}} e^{-i \hat{\phi}_T}\bigg) \rho_{n-1} \bigg( e^{i \hat{\phi}_T} \sqrt{\frac{\hat{p}_f}{\tilde{D}_+ \tilde{D}_-}} \tilde{D}_+\bigg),
 \end{split}
\end{align*}
where $\rho_n$ is the nuclear density matrix after $n$ cycles.

Rather than solve for the exact dynamics of the nuclear density
matrix--
still an intractably hard computational problem for any
reasonable number of nuclear spins--we instead adopt an approximate
solution to the problem using the P-representation for the density
matrix as an integral over products of spin coherent
states. 
From the thermal distribution, we choose such a spin coherent state
and evolve it, where we interpret expectation values $\mean{...}$ as
being taken in that state. The ensemble of such trajectories
represents the physical system \cite{Al-Hassanieh06}.

\providecommand{\ovec}[1]{\overset{\rightarrow}{#1}}
\newcommand{\boldnabla}{\bm{\nabla}}
\renewcommand{\Im}{\mathrm{Im}}
We organize this calculation by noting that the components of the
Landau-Zener model $(\hat \phi_S,\hat \phi_T, \hat p_f,\tilde{D}_{\pm})$ are only
functions of $\hat{\bm{L}}$ and $\hat{ \bm{R}}$. A spin coherent state is
entirely described by its expectation values $\vec{i}_{i\s} =
\mean{\vec{I}_{i\s}}$. For the $k$th spin on the left dot, we expand the
discrete time difference $\mean{\vec{I}_{kl}}_n - \mean{\vec{I}_{kl}}_{n-1}$ after $n$ and $n-1$ cycles in the small parameter $g_{kl}$, giving an evolution equation
\begin{equation} 
\frac{d\vec{i}_{kl}}{dt}
=g_{kl}\sum_{\mu=1}^3 P_{l,\mu} \mean{i[\partial_{g_{kl}} \hat L_\mu,\vec
  I_{kl}]}=g_{kl} \bm{P}_l \times \vec i_{kl}, \label{eq:EOM}
\end{equation}
where $\del_{g_{kl}}$ is the derivative with respect to $g_{kl}$ and 
\begin{align*}
  \bm P_l & =   \frac{1}{T} \left[\langle{1-\hat p_f}\rangle \langle{\boldnabla_l \hat \phi_S}\rangle + \langle\hat{p_f}\rangle \langle{\boldnabla_l \hat \phi_T}\rangle - \Im(\bm{\gamma}_l)\right],
\end{align*}
where $\boldnabla_l=(\partial_{L_x},\partial_{L_y},\partial_{L_z})$ are partial derivatives with respect to the variables $L_\mu$ and
\begin{align} \label{eqn:gamL}
  \bm\gamma_{l} = \mean{ \tilde{D}_{+} \frac{ \hat p_f}{\tilde{D}_- \tilde{D}_+} \boldnabla_l\tilde{D}_-},
\end{align}
and similarly for $\vec{i}_{kr}$, $\vec P_r$, and $\bm\gamma_r$, with $\vec{L}$ replaced by $\vec{R}$. The factorization of expectation values is a natural consequence of our spin-coherent state approximation, as it explicitly prevents entanglement between spins.
Thus we have an effective, semi-classical picture of nuclear spins precessing and being polarized by their interaction with the electron spin, integrated over one cycle.

\begin{figure}
\includegraphics[width=0.49 \textwidth]{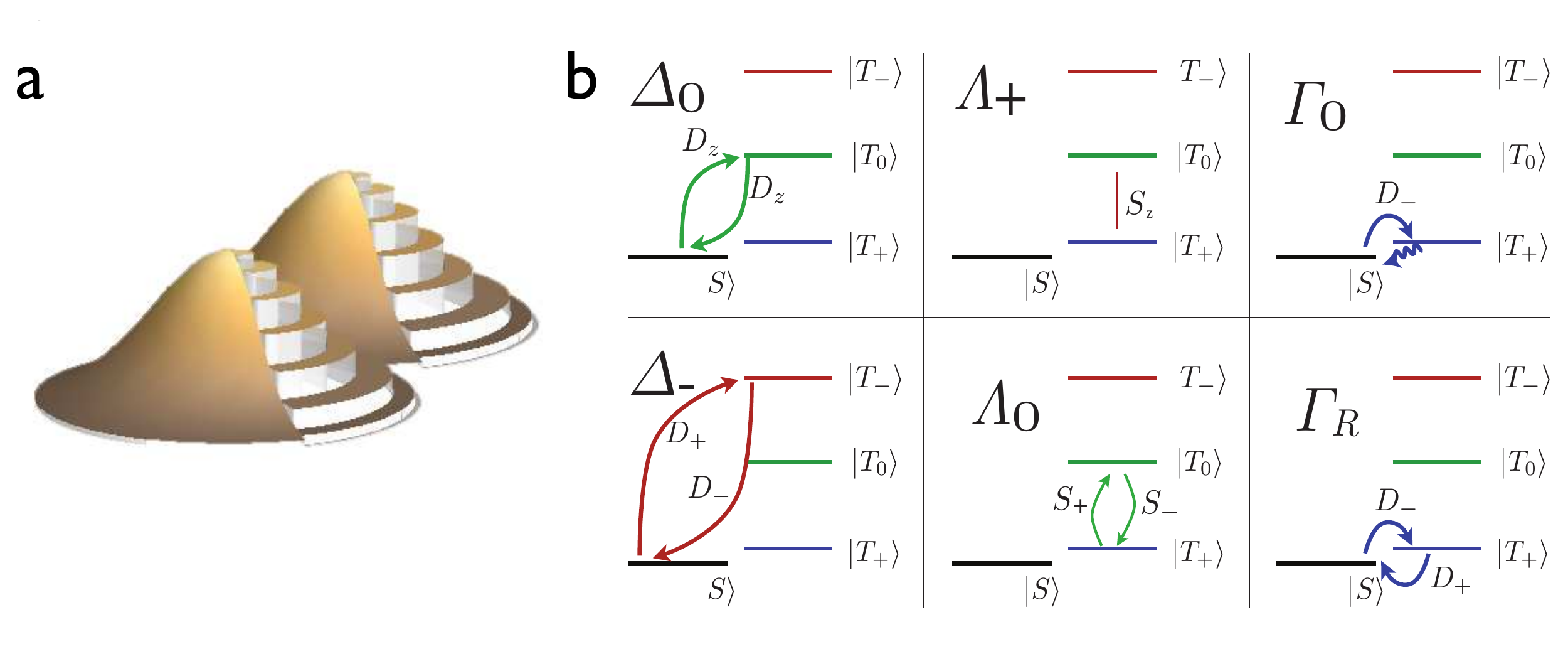}
\caption{\label{fig:cakedots} a) Independent Random Variable Annular Approximation (IRVAA) to the electron wavefunction in the double dot. b) Key processes contributing to Eq.~\ref{eq:PsPd}.}
\end{figure}

We approximate the electron wavefunction as a piecewise-flat function
with $\M$ levels, which we refer to as the annular approximation, as
illustrated in Fig.~\ref{fig:cakedots}a. Each annulus defines
$ \vec\I_{nd}=\sum_{k\in n} \vec i_{kd}$,
where the sum is over all nuclei with the same hyperfine coupling to
the electron. Since $g_k$ is identical for all $k\in n$, we can simply
replace $\vec i_{kd}$ with $\vec \I_{nd}$ in Eq.~\ref{eq:EOM}.  Furthermore,
$\I_n^2$ is a conserved quantity, so we can study the evolution of
$\M\ll N$ spins in a reduced Hilbert space.  The typical
size of $\I_n$ is $\sim \sqrt{N/M} \gg 1$, which allows us to
replace the spin-coherent states used above with semi-classical spins,
and makes taking expectation values straightforward: all quantum
operators can be replaced by their expectation values directly.   In the simulations presented in this work we took up to $\sim10^3$ distinct hyperfine coupling constants, which for such systems corresponds to $\sim 10^3$ spins per layer justifying the use of the semiclassical approximation \cite{Loss11}.  The
annular approximation should correctly describe the nuclear dynamics
for a time scale given by the inverse of the difference between the
$g_k$ of adjacent annuli.  

To illustrate, to \emph{first order} in $\m=\Bext^{-1}$, for $\s=l,r$, 
\begin{align}\label{eq:PsPd} 
\vec P_{\s}=&\pf\tred\left(\Lamw\zhat - \Lams\Sperpvec\right)+
\m\GamO\frac{\pf\Dz}{2\pi\omega^2}\zhat\times\Dvec \\\nonumber
&+\LamR \pf \boldnabla_{\s}\phi_{AD}\mp \Big[\GamO \frac{\beta^2}{4\pi v^2}\Im \left( \bm\gamma_{l}-\bm\gamma_{r}\right)
\\\nonumber
&+(1-\pf\tred/2)(\DelO\Dz\zhat + \Delm\Dperpvec)\Big]
\end{align}
where the top sign applies for $d=l$, $\Dperpvec=(D_x,D_y,0)$,
$\Sperpvec=(S_x,S_y,0)$, $\tred=1-2t_0/T$ gives the shift in the
location of the crossing, and $\DelO$, $\Delm$, $\Lamw$, $\Lams$,
$\LamR$, and $\GamO$ are constants depending on the details of the
pulse cycle (see below). We have replaced operators by their expectation values and removed the angle brackets since we are now in the semiclassical limit.  To leading order in $\m$, $\Im(\bm\gamma_{l}-\bm\gamma_{r})=2(\Dvec\times\zhat)\pf/\Dperp^2$. It
is clear from Eqs.~\ref{eq:EOM}-\ref{eq:PsPd} that all dynamics stop
in the zero states with $\Dvec=0$, consistent with the idea that true saturation of polarization requires that all components of $\Dvec$ be small.  We will focus on the stability of such states in various parameter regimes. The equations of motion in Ref.~\onlinecite{Gullans10} are found from Eq.\ \ref{eq:PsPd} by including only the lowest order in $\Omega\, T$ and $\Omega/\beta$, which is the limit of fast cycles and small spin flip probability per cycle, respectively.

First we outline the meanings of the parameters in the
model.  As indicated schematically in Fig.~\ref{fig:cakedots}b, the $\GamO$ term
originates in the hyperfine flip-flop, the $\DelO$ and $\Delm$ terms
are the off-resonant effects of coupling from the singlet state to the
$T_0$ and $T_-$ states, respectively, $\Lams$ comes from coupling
between the $\Tp$ and $T_0$ states, and $\Lamw$ comes from Knight
shifts due to occupation of the $\Tp$ state. To leading order in $\m$,
for a pulse sequence consisting of only the Landau-Zener sweep, with
instantaneous eject and reload, the parameters have values
\begin{align*}
\DelO&=\mean{\frac{2v^2 }{\J(t)}}_{\! \! \! \text{c}}\approx\m, &
\Delm&=\mean{\frac{v^2 }{\J(t)+\Bext}}_{\! \! \! \text{c}}\approx \m/4\\
\Lamw&=1/4, &
\Lams&=\m/4\\
\GamO&=\frac{2\pi v^2 f_c}{\beta^2}, &
\LamR&=f_c
\end{align*}
where $f_c=1/T$ is the cycle frequency and $\mean{.}_{\! \text{c}}$
indicates an average taken over a full cycle; these values can be
modified readily by changing the details of the pulse cycle, while
leaving the Landau-Zener portion unchanged.  In Appendix A we provide a reference for all parameters used in the simulations.

Equation \ref{eq:EOM} is a good approximation of the nuclear dynamics over a few DNP cycles because other nuclear processes are slow compared to a typical experimental cycle ($\sim$10-100 ns \cite{Reilly08a}).
However, the full DNP may last millions of cycles at which point these other nuclear processes become important.  Apart from Larmor precession, which is only relevant for the case of multiple nuclear species considered in Appendix C, nuclear quadrupole splittings and nuclear dipole-dipole interactions are the dominant processes.  They become relevant on a timescale of a few hundred microseconds in these systems \cite{Taylor07}.  We include them in our model phenomenologically by adding a fluctuating magnetic field $h_{kd}(t)$ in the $z$-direction at each site (the transverse terms are strongly suppressed by the external field), such that 
\be \label{eqn:EOMb}
\frac{d\bm{i}_{kd}}{dt} = g_{kd} \bm{P}_d \times \bm{i}_{kd} - \gamma_n\,h_{kd}\, \hat{z} \times \bm{i}_{kd}
\ee
where $\gamma_n$ is the nuclear gyromagnetic ratio.
We further assume that the this field can be treated as noise and characterized by a Gaussian, uncorrelated white noise spectrum
\be \label{eqn:eta}
\gamma_n^2 \langle h_{kd}^z(t) h_{k'd'}^z(t')\rangle_n= 2 \eta\, \delta(t-t') \delta_{kk'}\delta_{dd'}
\ee
where $\mean{\cdot}_n$ are averages over the noise \cite{Reilly08b}.   

%
%
%
%
%
%
%
%

\section{Results}

The polarization dynamics display three characteristic behaviors: growth of large difference fields, saturation in nuclear dark states defined by $D_\perp=0$, and preparation in zero states $\bm{D}=0$ which are global fixed points of the nuclear dynamics in the absence of noise.  In Ref. \onlinecite{Gullans10} this system was studied in a restricted model focusing on the case where noise was present.  Therein it was found that when the two dots have different hyperfine couplings the system generically grows large difference fields, while for identical dots, depending on parameters, the system is either unstable to the growth of large difference fields or saturates in dark states; however, the zero states were not found to be a relevant steady state in any parameter regime.  In the present work we focus on extending the results of Ref. \onlinecite{Gullans10} to a larger, more experimentally relevant, parameter regime by using equations of motion correct to second order in $m_0$ with a more complete model of the Landau-Zener sweep as described in the previous section.  In addition, we consider the nuclear dynamics in the absence of noise.  We also present the full analytical calculations which were omitted from Ref. \onlinecite{Gullans10}.  In all \emph{physical} parameter regimes we find qualitatively consistent results with Ref. \onlinecite{Gullans10}; however, for a limited, \emph{unphysical} parameter regime we do find solutions to the equations of motion in the absence of noise where the zero state is uniformly reached starting from a completely uncorrelated nuclear spin ensemble.  

\begin{figure}
\includegraphics[width=3.375in]{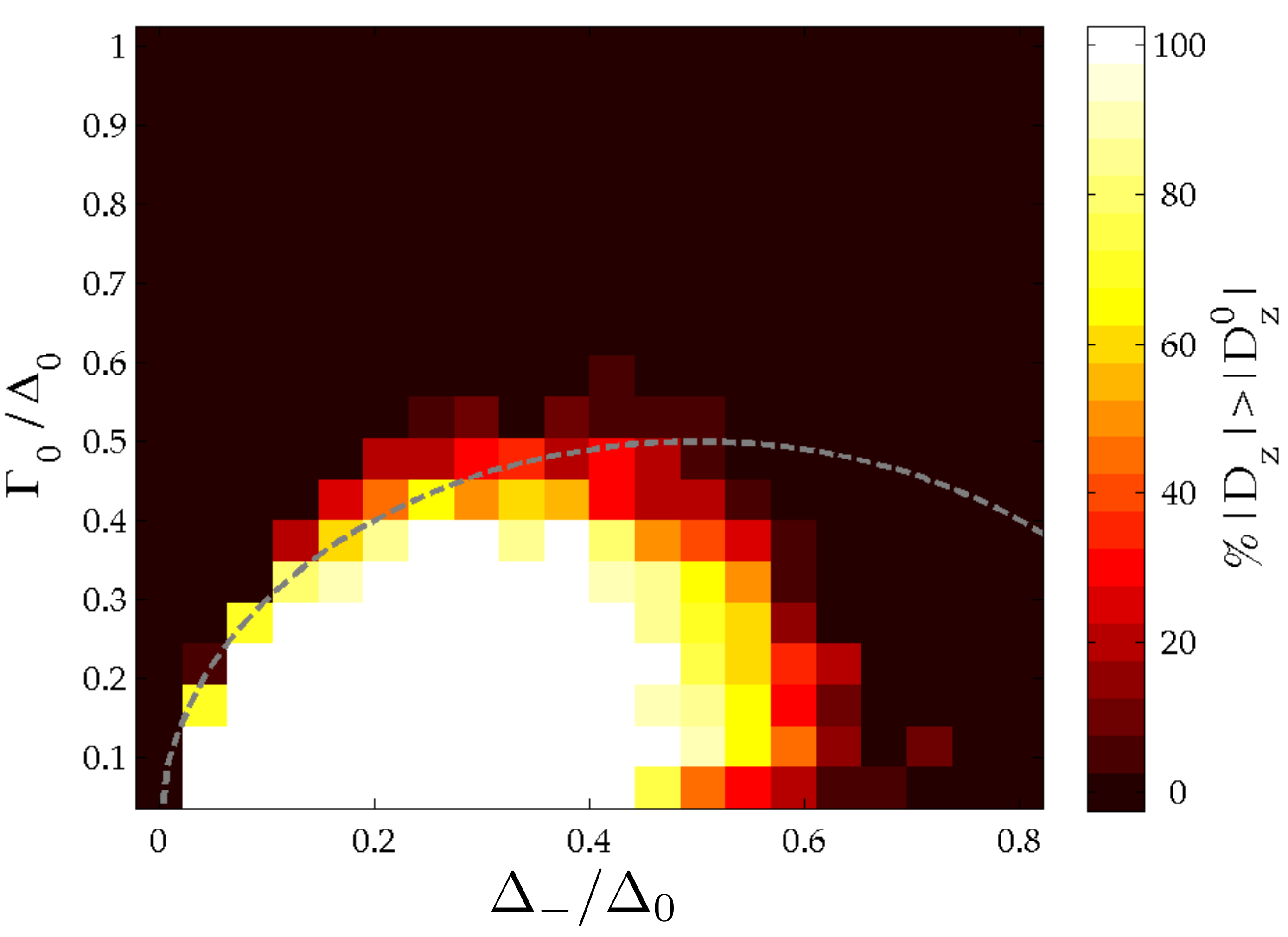}
\caption{\label{fig:phase1} Phase diagram for the simplified model presented in Ref.~\onlinecite{Gullans10}. At each value of parameters, twenty runs were started with $D_z=-2$, $S_z=-10$, and all other components chosen randomly according to the infinite temperature ensemble. The colorscale indicates how many of those runs ended with $|D_z|$ increased. The dark region is of saturation and the light region is of instability.  The dashed line shows the prediction of the simple model of Eq.~\ref{eqn:phaseBound}, which captures the phase boundary, especially at low $\Delta_-/\Delta_0$.  For parameters used, see Table I. }
\end{figure}

\begin{figure}
\includegraphics[width=3.375in]{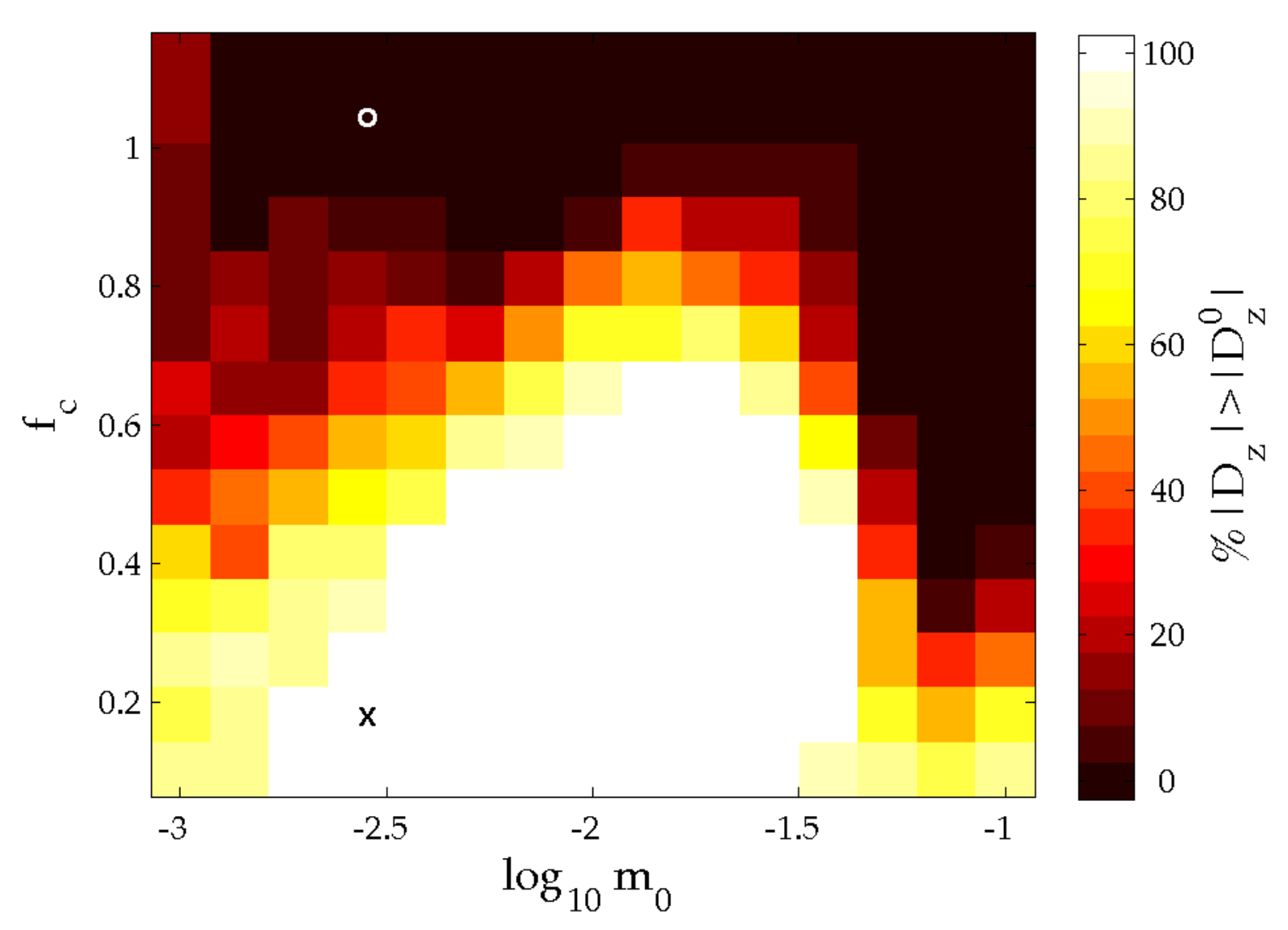}
\caption{\label{fig:phase2}  Phase diagram as in Fig.~\ref{fig:phase1}, except with varying external magnetic field and without any noise added. The parameters were scaled with $m_0$ as shown in Section III. There is a clear boundary between saturation at large $\Gamma_0$ and instability at lower values of $\Gamma_0$, with appropriately large values of $\Delta_0$ and $\Delta_-$. See Table I for parameters. The symbols 'x' and 'o' mark the parameters used for Fig.\ \ref{fig:ratio} below.} 
\end{figure}

The simulations shown below were performed with the equations of
motion correct to second order in $\m$ with $\psi_d(\vec r)$ a 2D Gaussian. Taking $v^2\approx1/2$,
we estimate that for experiments performed with $\Bext=10$~mT with
$T=25$~ns \cite{Reilly08a}, $\m\approx0.18$, $\GamO\approx0.20$, but
the $\Delta$ and $\Lambda$ terms depend on the rest of the cycle.
In each of the simulations, we choose initial magnitudes and directions of the spins $\vec\I_n$ by a procedure equivalent to choosing initial directions for each of the $N_n$ spin-3/2 nuclei in the n$^\text{th}$ annulus and evaluating $\vec\I_n=\sum_{k\in n} \vec i_k$ explicitly (see Appendix B).
The relationship between simulation time and laboratory time depends on the details of the pulse cycle, including pauses and reloads not considered explicitly here, but simulation time is roughly in units of $g_{max}^{-1}$, where $g_{max}\approx2\Omega^2/a_{hf}$ is the largest value of $g_k$, so $t=400$ is approximately 10~ms.

To organize our results we recall the phase diagram for identical dots and the simplified model derived in Ref.~\onlinecite{Gullans10} in which the only non-zero parameters are $\Delta_{0,-}$, $\Gamma_0$ and $\eta$, which corresponds to the limit of large magnetic fields and fast sweeps including nuclear spin noise.  To obtain the phase diagram we consider for each set of parameters whether the system supports self-consistent growth of $\abs\Dz$ starting from large values of $\abs\Dz$ and $\abs\Sz$.  This approach avoids complications with the metastability of zero states discussed later.   Such simulations produce the phase diagram in Fig.~2b of Ref. \onlinecite{Gullans10}, which is reproduced in Fig.~3 with the full data presented.
From this figure it is clear that we can separate the dynamics into two regimes depending on parameters.  For large ratios of $\Gamma_0/\Delta_0$, which corresponds to large magnetic fields or strong pumping the system quickly saturates with no growth of large difference fields.   For small ratios there is an instability towards large difference fields.  In the first section we explore the dynamics in the absence of noise for identical dots with all parameters included.  In the second section we include nuclear spin noise and asymmetry in the dot sizes.
  
\begin{figure*}[th]
\includegraphics[width=.8 \textwidth]{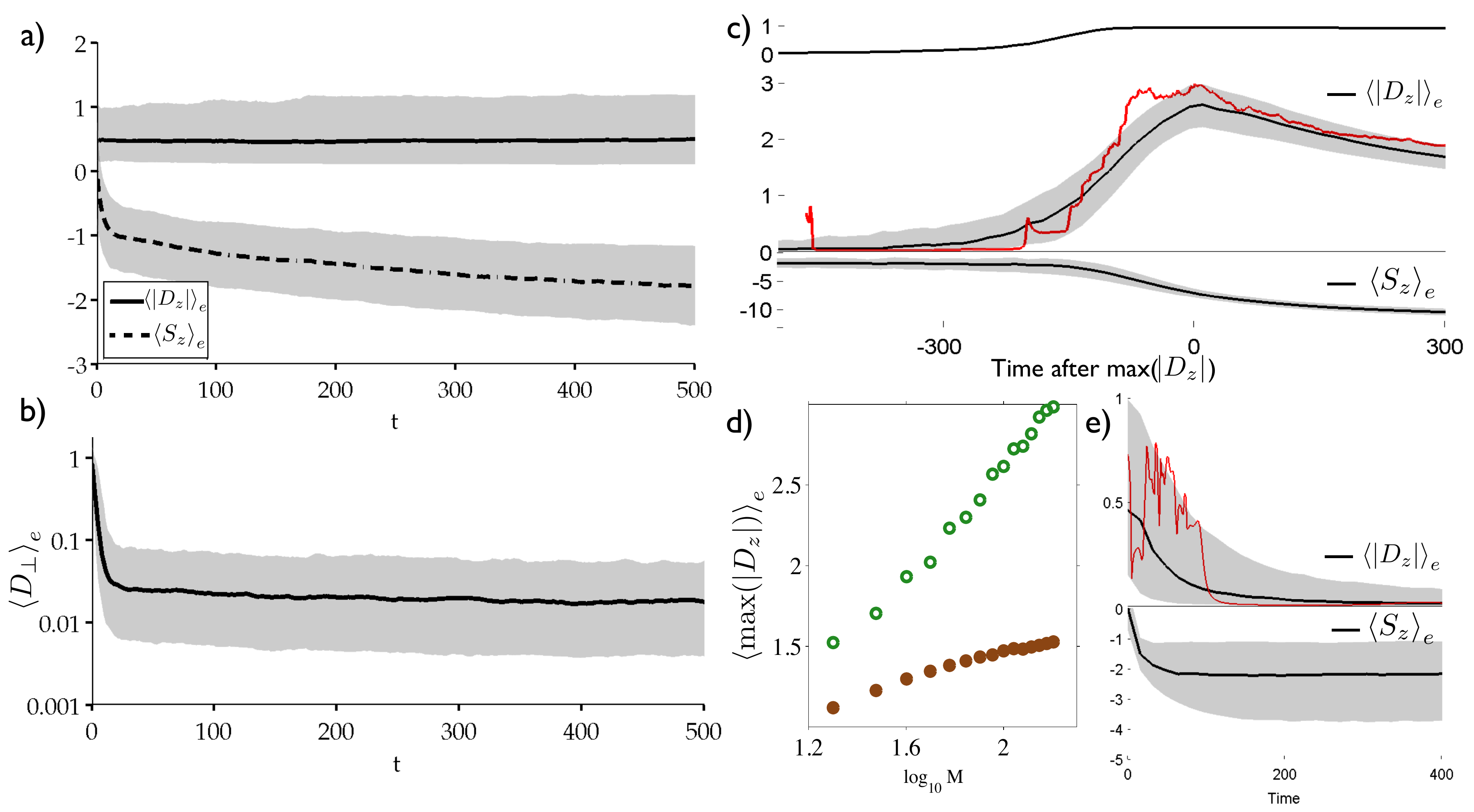}
\caption{\label{fig:traj} 
a-b) Simulations corresponding to the saturation region of the phase diagram.  The solid lines are the median values of $\abs\Dz$ (a), $\Sz$ (a) and $\Dperp$ (b) at each time step in an ensemble of 1000 trajectories. In all plots shaded regions show the $84^\text{th}$ and $16^\text{th}$ percentiles. 
c) Simulations showing growth of $\av{\abs{\Dz}}$ with the time shifted for each trajectory so that its maximum $\abs{D_z}$ occurs at time zero.  Bottom shows the median value of $S_z$ $\langle S_z\rangle_e$  at each time step in an ensemble of 1000 trajectories. In the middle is similar $\langle |D_z| \rangle_e$. Thin red line is a single trajectory. The curve at top shows the fraction of trajectories contributing to the ensemble at each time; this increases with time because some trajectories reach their maximum $D_z$ much later than others while the simulation time is fixed for each trajectory. 4.5\% of the trajectories, which do not show this peak in $\abs{\Dz}$, are not included. Approximately 10\% of the trajectories show behavior similar to that shown in the thin red line, where $\abs{\Dz}$ is reduced initially and then goes unstable to large $\abs{\Dz}$.
d) Mean of the maximum value of $\abs\Dz$ reached on each trajectory for the same parameters as in (c) (open circles) except $\M$ varied between 20 and 160, with 5000 trajectories per point. Closed circles show similar results with $\m=0.05$, $\tau=4$ and all other parameters scaled appropriately. The physical system has $\M\rightarrow N\approx 10^6$, so we interpret this as an instability to large $\abs\Dz$, which is supported by simulations including transverse noise (see section \ref{sec:NoiseIdentical}).
e) With different parameters, simulations showing reduction of $\av{\abs\Dz}$ plotted as in (c) without the time shift.  For these parameters, the trajectories have $|D_z|\rightarrow0$ quickly, without time for strong polarization.
%
%
}
\end{figure*}

\subsection{Noise Free Nuclear Spins}

From the general arguments given in the introduction it is clear that when the dots have different hyperfine couplings the system naturally grows a large difference field.  Furthermore, in Ref. \onlinecite{Gullans10} it was shown that even identical dots display similar behavior in the presence of nuclear spin noise.  Although a complete theory of the polarization dynamics must take into account nuclear dipole-dipole interactions (which we approximate by noise), it is still useful study the dynamics in the absence of noise.  Such analysis is relevant especially for short times and helps gain an understanding of the role of the coherent nuclear dynamics.  Therefore, in this section we analyze the case of identical dots in the absence of noise.     We begin by deriving a phase diagram analogous to the one obtained in the presence of noise except we now look in the space of the experimentally accessible  parameters cycle rate $f_c$ and inverse magnetic field $m_0$.  The results are shown in Fig.~\ref{fig:phase2} where we see the same qualitative behavior as shown in Fig.~\ref{fig:phase1}.  However, the dynamics are much richer than indicated by this simple phase diagram. In the following subsections we give examples of what happens to a nuclear spin ensemble starting from equilibrium for different parameters and regions of the phase diagram. 

Before proceeding, however, we note that in the absence of noise the inhomogeneity of the electron wavefunction plays a crucial role.  This is because weak inhomogeneity is equivalent to choosing the number of annuli $\M$ to be small and in this case the system moves rapidly to its maximally polarized state, with $\vec\I_n\approx-\I_n\zhat$ for all $n$.  Dynamics completely cease in this state, as can clearly be seen from Eq. \ref{eq:EOM}, despite the fact that this state does not correspond to all of the nuclei being polarized, which would also require $\I_n=3N_n/2$.
On the other hand, for strong inhomogeneity, or large $\M$, when the system is not fully polarized other terms in $\bm{P}_d$ compete with the polarization saturation and sustain the dynamics. 
\cite{Christ07}

\subsubsection{Polarization Saturation}

When the magnetic field is large or the cycle rate is fast (i.e., $\DelO\ll \GamO$), the system rapidly moves toward dark states (i.e., states with $\Dperp=0$), sending $\pf\rightarrow0$ without statistical change in the distribution of $\Dz$, as shown in Fig.~\ref{fig:traj}a. This limit is additionally characterized by only a small change in nuclear polarization as seen in Fig.~\ref{fig:traj}b. When the effects of the $\ket{s}$-$\ket{\TO}$ coupling are important (i.e., $\DelO\approx\GamO$), the $\DelO$ term in Eq.~\ref{eq:PsPd} causes $\Dperp$ to increase, ``rebrightening'' the $\Dperp\approx0$ dark states and allowing dynamics to continue. Coupling from the singlet to the $T_0$ state is an essential ingredient in all of the effects discussed below. When $\DelO$ is significant, dynamics only stop near zero states with $\Dvec=0$.


\subsubsection{Growth of difference fields}
Second, we observe the growth of large Overhauser fields.
We consider a prototypical pulse sequence motivated by experiments with moderate/large magnetic field, $\m=0.01$ 
In this case, over 95\% of the trajectories display a growth in $\abs{\Dz}$, as shown in Fig.~\ref{fig:traj}c. We observe this behavior over a range of experimentally accessible magnetic fields and cycle frequencies.
This increase in $\abs{\Dz}$ indicates that the spin flips are occurring predominantly in one dot. We interpret these results as showing a continuing increase of $\abs{\Dz}$, where the peak of $\abs{\Dz(t)}$ is an artifact of the annular approximation. Near the peak, many of the annular spins artificially reach their maximal polarization, at which point they should be broken into more annuli. Similar trajectories with different $\M$ show the maximum value of $\abs\Dz$ increasing with $\M$ (Fig.~\ref{fig:traj}d). 
The physical cause of this increase in $\abs\Dz$ is not clear, but it is associated with both $\DelO/\GamO$ and $\Lamw/\GamO$ being sufficiently large. When  nuclear spin noise is  included, the growth in $\av{\abs\Dz}_e$ continues \cite{Gullans10}.  This could be the same phenomenon as seen in Ref.~\onlinecite{Foletti09}, with transverse dephasing helping to produce the large $\abs\Dz\approx B_\text{ext}$ of that work, though unequal dot sizes could also produce that effect \cite{Gullans10}. 

\subsubsection{Zero states}
For moderate to small magnetic fields, when $\DelO\approx \GamO$, two different characteristic behaviors of particular note are observed.  First, in the physical parameter regimes, which do not display general motion to zero states, the zero states are still important for the dynamics as they are a metastable state. That is, many trajectories spend a long time with $\abs\Dz$ near zero before escaping away to large $\abs\Dz$.  This phenomenon is shown in the individual trajectory (thin red line) of Fig.~\ref{fig:traj}c. 

Second, for parameters in our model which are not experimentally accessible there is a mechanism that gives rise to attraction towards zero states.  This is illustrated in Fig.~\ref{fig:traj}e, where we show an ensemble of trajectories in which $\Dvec$ rapidly reduces toward zero. 
For the parameters of Fig.~\ref{fig:traj}e, the standard deviation of $\Dz$ was reduced by a factor of 28. We remark that as $\Dvec\rightarrow0$, the singlet state ceases mixing with the triplets and nuclear spin dynamics stop. Until something (outside this model, such as nuclear dipole-dipole coupling) restores $\Dvec$, the polarization process is shut off, limiting the total nuclear polarization that can build up.  While not shown in Fig.~\ref{fig:traj}e, we observe a dramatic reduction of the total $|D|$, not just $D_z$, consistent with this qualitative observation.  However, because we have not observed this phenomenon in any physical parameter regimes we shall not study it further.

\subsubsection{Crossover}

\begin{figure}
\includegraphics[width=3.375in]{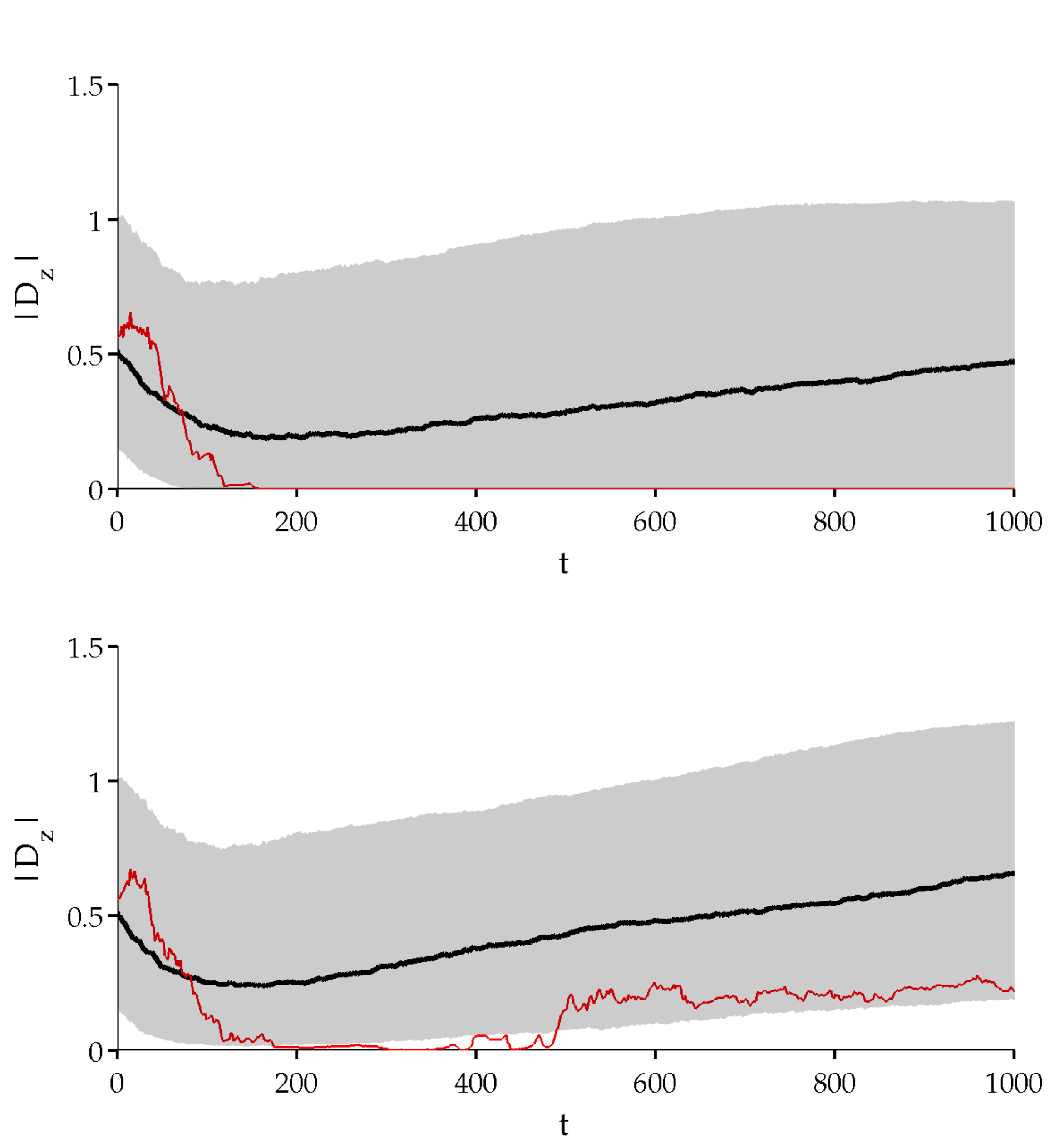}
\caption{a) 1000 trajectories were run with initial conditions chosen from the thermal distribution with no noise. The mean value of $\abs{D_z}$ is shown in black, and the the gray region enclosing 67\% of the trajectories. A single trajectory is shown in the thin red line. For parameters, see Table
1.  These parameters are not represented in the phase diagram since they have very large $\Lambda_+$.  For these parameters, many trajectories are attracted near $D=0$, as in the single trajectory shown, for extended periods of time. b) Trajectories were begun from identical configurations as in a, this time with noise added. With noise included, the metastability of the zero state is removed, and the gray region is now bounded away from zero.  \label{fig:crossover}}
\end{figure}

For many choices of parameters, we find both trajectories in which $\Dz\rightarrow0$ and $\abs\Dz$ remains large, depending on initial conditions, as shown in Fig.\ \ref{fig:crossover}a. Note that when we add a small amount of transverse dephasing to these trajectories, as shown in Fig.\ \ref{fig:crossover}b, the median value of $\abs\Dz$ does not markedly change, but there are no longer trajectories with $\Dz\rightarrow0$; the noise apparently disrupts the fragile attraction toward $\abs\Dz\rightarrow0$. 
Simulations performed with parameters intended to approximate experiments \cite{Reilly08a,Foletti09} are in this crossover regime.

\subsubsection{Stability of zero states}

We now investigate  more carefully the stability of the zero states.  Near the zero state the EOM are greatly simplified because many of the terms in $\bm{P}_d$ arise from perturbative processes involving multiple applications of $\bm{D}$.  Keeping only the terms linear in $\bm{D}$ and working to first order in $m_0$ we can write
\begin{align}
 \dot{D}_+&=\big(\Gamma_0+i \Delta_-) S^*_z D_+ + (\Gamma_0 m_0 S_z^* S_+ - i \Delta_0 S_+^*)D_z \\
 \dot{D}_z&=- \textrm{Re}\big[ (\Gamma_0 +i \Delta_-)D_+ S_-^* \big] - \Gamma_0 m_0 \bm{S}_\perp \cdot \bm{S}_\perp^* D_z , \label{eqn:Dzdot}
 \end{align}
 where we have introduced the variable $\bm{S}^*= \sum_{kd}g_{kd}^2 \bm{I}_{kd}/2$.  Because $d\bm{S}/ d t,d \bm{S}^* / d t \sim O(D)$, we can neglect the time dependence of $\bm{S}$ and $\bm{S}^*$ in the EOM for $\bm{D}$ near the zero state.  After a long time the system becomes polarized so that $S_z^* \ll 0$, this allows us to adiabatically eliminate $D_+$ to obtain
 \begin{align}
 D_+& = \frac{- i \Delta_0 S_+^*+ m_0 \Gamma_0 S_z^* S_+ }{(\Gamma_0 + i \Delta_-)\abs{S_z^*}} D_z + O(D^2)\\
 \dot{D}_z& = 0 + O(D^2)
 \end{align}
This linear stability analysis gives no conclusion about the stability of the zeros states.  This result implies that within this model the stability of the zero state is only determined at higher order.  This is a little surprising because at first glance Eq.~\ref{eqn:Dzdot} appears to have an attractive force towards $D_z=0$.  This arises from the same mechanism described in Ref. \onlinecite{Stopa10}; however, a more careful treatment reveals that this effect actually cancels.   Our simulations indicate that the nonlinear corrections make the zero state repulsive in the experimentally relevant parameter regimes.  When we include the nuclear spin noise we shall show analytically that the system is repelled from the zero states.    


\subsection{Effect of Nuclear Spin Noise}
\subsubsection{Unequal Dots}

\begin{figure}
\includegraphics[width=3.in]{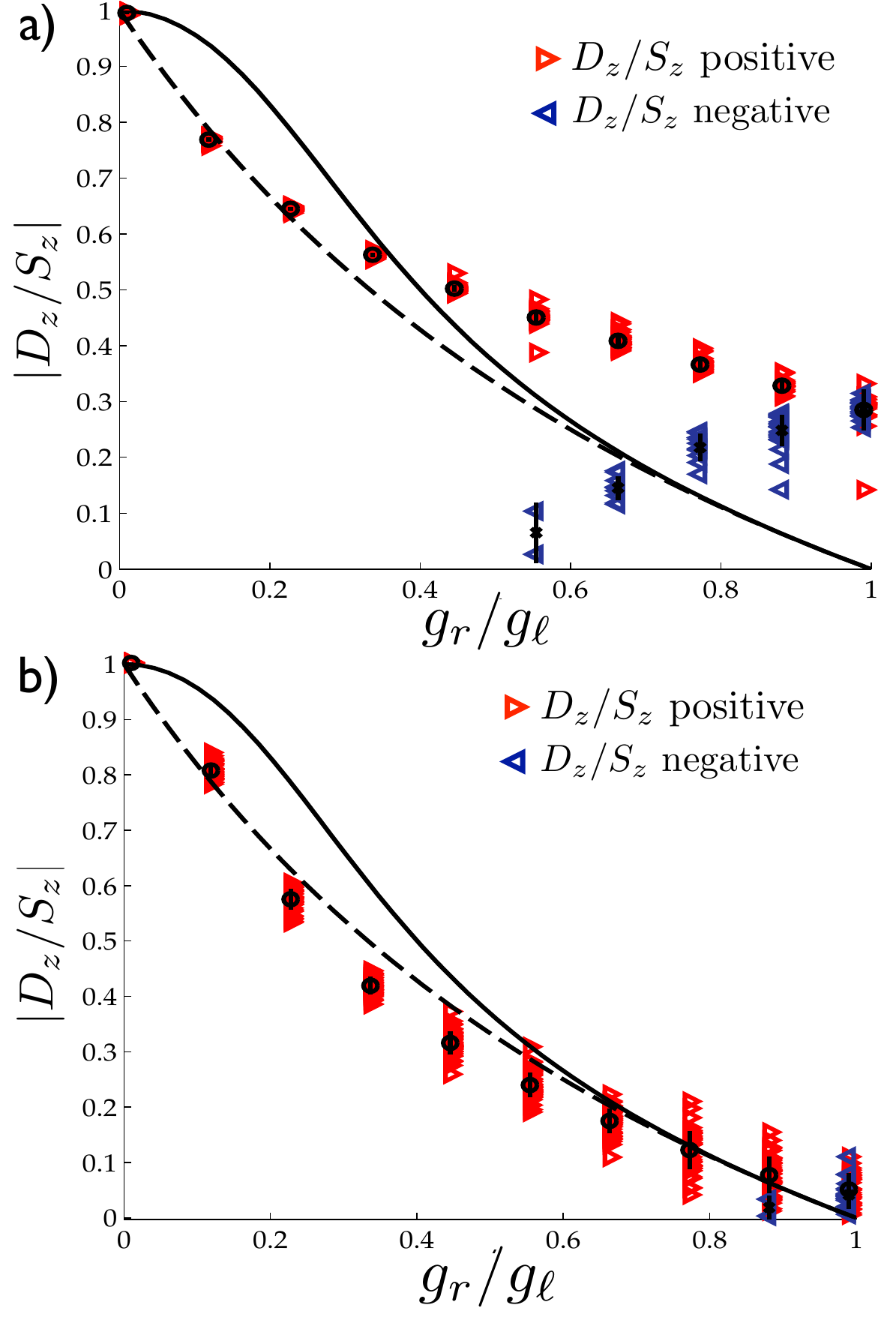}
\caption{\label{fig:ratio} a) Asymptotic value of $\abs{D_z/S_z}$ as a function of dot asymmetry with parameters chosen as in the location marked with an x in Fig.~\ref{fig:phase2}, strongly in the instability regime. The horizontal access corresponds to the left dot decreasing in size from right to left, which, by our simple argument, should result in a positive ratio of $D_z/S_z$.  Trajectories which show the opposite sign indicate a competition with the coherent instability mechansim. For each value of dot asymmetry $R$, we initialized fifty runs in a single initial spin configuration chosen from the thermal distribution (with $D_z=-0.72$ and $S_z=-1.57$). We plot the asymptotic value of $\Dz/\Sz$. The runs that ended with $D_z/S_z$ greater (less) than 0 shown are shown as red (blue) points. The circles (crosses) indicate the mean value of the red (blue) points, with error bars showing the standard deviations. The solid and dashed lines are given by Eq.~\ref{eqn:ratioPrec} and Eq.~\ref{eqn:ratio}, respectively.  b) As in (a), with parameters chosen in the location marked with an o in Fig.\ \ref{fig:phase2}, strongly in the saturation regime.  Here the sign of the ratio $D_z/S_z$ follows what is expected from the natural asymmetry.  } 
\end{figure}

Our results that zero states are unstable to
the growth of large difference fields in the presence of asymmetry in the
 size of the dots and nuclear noise can be 
be understood in the following heuristic picture first given in Ref.\ \onlinecite{Gullans10}.  We assume the nuclear spins have equal spin flip rates on the two dots, which is borne out by the analytical and numerical calculations presented below. Then the build-up of the total Overhauser field $S_z$ is proportional to $-(g_\ell + g_r)$, where $g_{\ell(r)}$ are the effective hyperfine interactions on the left (right) dot and the negative sign arise because nuclear spins are flipped down in the experimental cycles.  Similarly $D_z$ grows as $-(g_\ell - g_r)$ so that the ratio 
\be \label{eqn:ratio}
D_z/S_z \to (g_\ell - g_r)/(g_\ell+g_r).
\ee   

 In this section we demonstrate a similar result within our full model.  We assume homogeneous coupling and work  in the high field, large $J$, limit where we can set $\Delta_0=\Delta_- =0$ in $\vec{P}_d$.  
The local noise processes included in Eq.~\ref{eqn:EOMb} give rise to a mean decay of the collective nuclear spin variables and associated fluctuations $\mathcal{F}_{\ell(r)}$, for $\dot{L}_+(\dot{R}_+)$, defined by $\mean{\mathcal{F}_d(t)\, \mathcal{F}_{d'}^*(t')}_n = 2 \Omega_d^2 \, \delta_{dd'} \delta(t-t')$.  The semiclassical EOM for the nuclear spins reduce to
\begin{align}
\dot{L}_+&= g_\ell \Gamma_0 \, L_z (L_+ - R_+)/2 - \eta\, L_+ + \sqrt{2 \eta} \, \mathcal{F}_\ell,\label{eqn:incohLp}\\
\dot{L}_z & = - \frac{g_\ell}{2} \Gamma_0\,  \big(L_\perp^2 -  \vec{R}_\perp \cdot \vec{L}_\perp \big),\label{eqn:incohLz}
\end{align}
and similarly for $\vec{R}$, where $\eta$ is defined in Eq.~\ref{eqn:eta}.   From Eq.~\ref{eqn:incohLp}, we see that if we start in a zero state, $\mathcal{F}_d$ will produce a fluctuation in $\Dperp$, and the contribution to $\dot{L}_z$ of the form $- g_\ell \Gamma_0 L_\perp^2$ results, in the long time limit, in $L_z \ll -1$ and similarly for $R_z$.  Thus, $\vert\dot{L}_z/L_z \vert\ll 1$ and we can treat $L_z$, $R_z$ as static to find $\mean{L_\perp^2}_n$, $\mean{R_\perp^2}_n$ and $\mean{\vec{L}_\perp \cdot \vec{R}_\perp }_n$, which allow us to find the slow evolution of $L_z$, $R_z$. 

In particular, assuming $L_z$, $R_z$ are constant we can write the closed set of equations for $L_+$ and $R_+$
\begin{equation}
\begin{split}
\left( \begin{array}{c} 
\dot{L}_+ \\ 
\dot{R}_+
\end{array} \right)&= \frac{\Gamma_0}{2} \left(
\begin{array}{c c} g_\ell L_z & -g_\ell L_z \\
-g_r R_z & g_r R_z 
\end{array} 
\right)
\left( \begin{array}{c} L_+ \\ R_+\end{array} \right)\\
& - \eta \left( \begin{array}{c} L_+ \\ R_+\end{array} \right)
  + \sqrt{2 \eta} \left( \begin{array}{c} \mathcal{F}_\ell \\ \mathcal{F}_r\end{array} \right)
\end{split}
\end{equation}
Introducing the variables
\be
\left( \begin{array}{c} 
\tilde{S}_+ \\ 
D_+
\end{array} \right)= \frac{1}{2} \left(
\begin{array}{c c}-1 & - \frac{g_\ell L_z}{g_r  R_z} \\
1 & -1
\end{array} 
\right)
\left( \begin{array}{c} L_+ \\ R_+\end{array} \right)
\ee
we find
\begin{align}
\tilde{S}_+(t) & =-\sqrt{\frac{\eta}{2}} \int_{- \infty}^t dt' e^{-\eta (t-t')} \big(\mathcal{F}_\ell + \frac{g_\ell L_z}{g_r R_z} \mathcal{F}_r \big), \\
D_+(t)&= \sqrt{\frac{\eta}{2}} \int_{-\infty}^t dt' e^{-(\eta+\gamma_S)(t-t')} (\mathcal{F}_\ell - \mathcal{F}_r)
\end{align}
here $\gamma_S= - \Gamma_0 (g_\ell L_z+g_r R_z)/2>0$.
We can use this solution to calculate $\mean{L_\perp^2}_n,~\mean{R_\perp^2}_n$, and $\mean{\bm{L}_\perp\cdot\bm{R}_\perp}_n$. For example to lowest order in $1/L_z$, $1/R_z$
\be
\begin{split}
\mean{{L}_\perp^2}_n&= \frac{4 \eta / \bar{g}}{(1+ p)^2}\\
\times& \bigg( \frac{g_\ell+g_r p^2}{2 \eta} + \frac{(g_\ell+g_r) p^2}{2 \gamma_S} + \frac{2 p (g_\ell - g_r p)}{\gamma_S}\bigg),
\end{split}
\ee
where we have defined  $p=g_\ell L_z/g_r R_z$, $\bar{g}=(g_\ell+g_r)/2$ and used the fact that $\Omega_d^2=g_d /\bar{g}$ in our units.

Inserting this solution into the EOM for $D_z$, $S_z$ gives reduced EOM for the slow, noise-averaged evolution of $D_z$ and $S_z$.  After some straightforward manipulations we arrive at
\be
\left( \begin{array}{c} 
\dot{S}_z \\ 
\dot{D}_z
\end{array} \right)= \frac{g_\ell\, \eta}{2 \bar{g}} \frac{g_\ell\, g_r}{\abs{S_2^z}^2}~E \left( \begin{array}{c} S_z \\ D_z\end{array} \right)
\ee
where $S_2^z = (g_\ell L_z+g_r R_z)/2= -\gamma_S/\Gamma_0$ and 
\begin{align}
E=  \frac{1}{4R}\left(\begin{array}{c c} 
(1+R)(1-R^2) & (1-R)^3 \\
(1-R)(1+R)^2 & -(1+R)(1-R^2)
 \end{array} \right)
 \end{align}
 and $R=g_r/g_\ell$.
 After rescaling time to 
 \be
 \tau = \int_{0}^t dt'\,  \frac{g_\ell\, \eta}{ \bar{g}} \frac{g_\ell\, g_r}{\abs{S_2^z(t')}^2}
\ee
this becomes a purely linear system characterized by the matrix $E$.  For all $R>0$, this matrix has one positive and one negative eigenvalue; thus, it has one growing mode and one decaying mode.  In the long time limit, both $S_z$ and $D_z$ will be proportional to their overlap with the growing mode.  Thus $D_z/S_z$ approaches a constant, which is easily found from $E$ as  
\be \label{eqn:ratioPrec}
\frac{D_z}{S_z}\to  \frac{1-R^2}{2R + \sqrt{4R^2+(1-R)^4}}.
\ee

In Fig.~\ref{fig:ratio} we compare this result and Eq.~\ref{eqn:ratio} to the full numerics including all the parameters.  The horizontal access corresponds to the left dot decreasing in size from right to left, since $D_z/S_z \sim (g_\ell - g_r)/(g_\ell+g_r)$ according to our simple argument we expect this to result in a positive ratio of $D_z/S_z$.  In Fig.~7a, however, we see that  for small asymmetry $g_r/g_\ell >0.5$, many trajectories have the opposite sign indicates that in this regime the coherent instability mechanism (which does not prefer either sign) competes with the natural asymmetry.  For larger asymmetries $g_r/g_\ell <0.5$ all trajectories are seen to follow the direction of the natural asymmetry.  
%
%
Fig.~\ref{fig:ratio}b shows the same simulations performed in the saturation regime.  As there is no coherent instability mechanism competing with the dot  asymmetry, the sign of $D_z$ is determined by the asymmetry in all but the  most symmetric dots. $\Dz/\Sz$ is in good agreement with the simple  prediction given by Eq.~\ref{eqn:ratio} and Eq.~\ref{eqn:ratioPrec}.

%

\subsubsection{Identical Dots \label{sec:NoiseIdentical}}

For identical dots the arguments given in the previous subsection break down; however, we shall now show that for certain parameters there still exists a mechanism for self-consistent growth of $\abs{D_z}$.
Growth of $\abs\Dz$ requires nonzero $\Dperp$.   For intermediate field and exchange, the $\Delta_{0,-}$ contributions to $\bm{P}_d$ become comparable to the $\Gamma_0$ term.  In particular, the $\Delta_0 D_z \hat{z}$ term acts as a source term for $D_\perp$ (see Eq.~\ref{eqn:idEOM}).  Consequently, for weak enough noise $D_\perp$ will only be appreciable when $\abs{\Delta_0 D_z /\Gamma_0 S_z}$ is appreciable, which provides a self-consistency condition for the continued growth of $D_z$.

\begin{figure}
\includegraphics[width=3.375in]{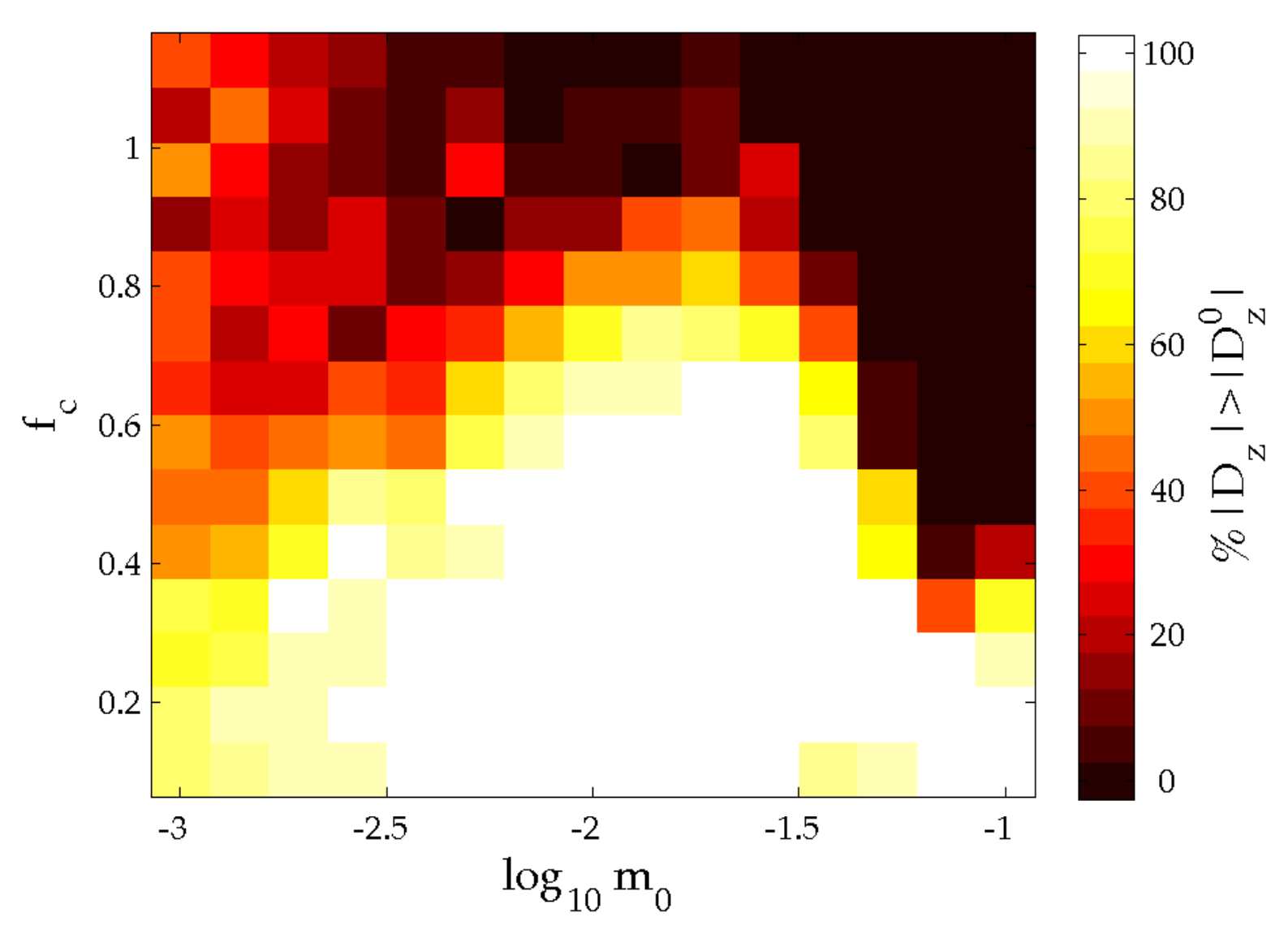}
\caption{\label{fig:phase3} Phase diagram as in Fig.~\ref{fig:phase2} except with noise added. The phase diagram is nearly identical. See Table I for parameters.   } 
\end{figure}

These properties of identical dots can be seen analytically in the
following limiting case: we assume a wave function where the coupling
takes two values, $g_1  \gg g_2, \eta$ and that initially $-g_2 S_z \gg g_1 \abs{D_z} \gg g_1$, $S_\perp \sim 1$ and $D_\perp  \sim D_z/S_z \ll 1$.  We denote the \emph{total} angular momentum of nuclear spins in dot $d$ with coupling constant $g_k$ by $\bm{J}_{kd}$  and assume $ J_{1d}^\perp \sim J_{2d}^\perp \sim J_{2d}^z \ll J_{1d}^z $ so that the majority of the polarization resides in the strongly coupled spins.  We can write a closed set of equations for the evolution of $\bm{D}$ and $\bm{S}$
\begin{align*}
\dot{D}_+&=g_1 i \tilde{\Delta}_- S_z D_+ - g_1 i \Delta_0 D_z S_+ \\
 +& g_2  \, \delta\, i \, \Delta_0 D_z (J_{2\ell}^+ + J_{2r}^+)/2 
 - g_2 \, \delta \, i \tilde{\Delta}_-\, D_+ (J_{2\ell}^z+J_{ 2r}^z)/2 ,\\
\dot{S}_+&= -g_1 i (\Delta_0 - \tilde{\Delta}_-) D_z D_+ +  g_2 \, \delta\,i\, \Delta_0 D_z (J_{2\ell}^+- J_{2r}^+)/2 \\
&-  g_2 \, \delta \, i\, \tilde{\Delta}_- D_+ (J_{2 \ell}^z - J_{2r}^z)/2, \\
\dot{D}_z&= g_1\, \textrm{Im}\big[ \tilde{\Delta}_- D_+ S_- \big] - g_2 \, \delta\, \textrm{Im}\big[ \tilde{\Delta}_- D_+ (J_{2\ell}^-+J_{2 r}^-)/2 \big], \\
\dot{S}_z&= -g_1 \Gamma_0 D_\perp^2 - g_2 \, \delta\, \textrm{Im}\big[ \tilde{\Delta}_- D_+ (J_{2\ell}^- - 
J_{2r}^-)/2 \big],\\
\dot{J}_{2d}^+&=\pm g_2 i \Delta_0 D_z J_{2d}^+ \mp g_2 i \tilde{\Delta}_- D_+ J_{2d}^z - \eta J_{2d}^+ + f_d  ,\\
\dot{J}_{2d}^z&=\pm g_2 \, \textrm{Im}\big[ \tilde{\Delta}_- D_+ J_{2d}^- \big],
\end{align*}
where the top sign is for $d=\ell$, $\tilde{\Delta}_- \equiv \Delta_- - i \Gamma_0$, $\delta\equiv g_1- g_2$, $f_d$ is a gaussian, white noise process derived analogously to $\mathcal{F}_d$ such that $\mean{f_d f_d^*}_n=2 \eta \sigma^2$, and we have neglected to write the noise terms in the EOM for $D_+$ and $S_+$ because we have assumed they are higher order.
Furthermore, we can neglect all terms proportional to $g_2 D_+ J_{2d}^\mu$ because these are second order.  This leads to the somewhat simpler set of equations
\begin{align} \label{eqn:idEOM}
\dot{D}_+&=g_1 i \tilde{\Delta}_- S_z D_+ - g_1 i \Delta_0 D_z S_+ \\
&+ g_2  \, \delta\, i \, \Delta_0 D_z (J_{2\ell}^+ + J_{2r}^+)/2, \nonumber\\
\dot{S}_+&= -g_1 i (\Delta_0 - \tilde{\Delta}_-) D_z D_+ \\
&+  g_2 \, \delta\,i\, \Delta_0 D_z (J_{2\ell}^+- J_{2r}^+)/2, \nonumber\\
\dot{J}_{2d}^+&=\pm g_2 i \Delta_0 D_z J_{2d}^+ - \eta J_{2d}^+ + f_d  ,\\
\dot{D}_z&= g_1\, \textrm{Im}\big[ \tilde{\Delta}_- D_+ S_- \big] \\
\dot{S}_z&= -g_1 \Gamma_0 D_\perp^2, 
\end{align}
These equations can be solved perturbatively in $1/S_z$,$1/D_z$ by the same method as in the previous section.  The only difference in the structure of the two problems is that in this case the source terms for $D_+$ and $S_+$ are proportional to $J_{2d}^+$ instead of white noise; as a result we have to take into account the coherent evolution of the source term. We can expand the resulting EOM for $D_z$ in $g_1 D_z/g_2 S_z$ to find the noise-averaged equation
\be
\begin{split} \label{eqn:phaseBound}
\dot{D}_z&=-  g_1 \Gamma_0 2 \, \delta^2 \sigma^2 \bigg( \frac{\Delta_0^2}{\Gamma_0^2+\Delta_-^2} \bigg)
\\
& \times\frac{\big(\Gamma_0^2+\Delta_-^2- \Delta_0 \Delta_-\big)}{\Gamma_0^2+\Delta_-^2} \frac{g_1}{g_2} \bigg(\frac{D_z}{\abs{S_z}}\bigg)^3
\end{split}
\ee
from which we see that the sign of $\Gamma_0^2+\Delta_-^2- \Delta_0 \Delta_-$ determines whether or not there is continued growth of $D_z$.  Note that the perturbation theory breaks down as $g_2\to 0$. This reflects the importance of including the coherent evolution of $J_{2d}^+$ in solving for the dynamics. Without $g_2$, we would have found $\dot{D}_z=0$.  This phase boundary is shown as the dashed line in Fig.~\ref{fig:phase1}.  In Fig.~\ref{fig:phase3} we show the phase diagram as a function of cycle frequency and inverse magnetic field, where we see qualitatively the same behavior as Fig.~\ref{fig:phase2}.

%
\section{Relevance to other  Central Spin Systems}
Although this work has focused on lateral double quantum dots in GaAs,  the methods, and some of the results, can be applied to vertical double dots \cite{Tarucha11}, InAs quantum dots \cite{Steel12,Imamoglu12}, silicon based quantum dots \cite{Maune12}, and NV-centers in diamond \cite{Childress06}.  A few important differences for these other central spin systems are that the sign of the electron $g$-factor may be positive (compared to its negative sign in GaAs) and the spin-orbit coupling can be much larger in other systems than it is in GaAs \cite{Rudner11}.
The results presented in the paper are not dependent on the sign of the $g$-factor.  Changing the sign would reverse the direction of the nuclear polarization from negative to positive, but all of our analysis would carry through essentially unchanged.  The competition between spin-orbit coupling and DNP is more dramatic and can have a qualitative effect on the polarization dynamics for large spin-orbit coupling \cite{Rudner10}.

\section{Conclusions}

We have shown that dynamic nuclear polarization experiments in double quantum dots give rise to a rich set of phenomena.  We find that after many thousands of nuclear spin pumping cycles, corresponding to experimental timescales of several hundred microseconds, the total nuclear polarization is driven  to $10-30 \%$ of full polarization.  The polarization is aligned opposite the magnetic field as opposed to the thermal polarization.  In addition to this large polarization, we find
the competition between polarization, noise processes and coherent evolution mediated by the electrons allows one to carefully control the final nuclear spin state in the two dots.    We have developed detailed numerical and analytical methods to theoretically describe such dynamics;  however, our analysis is semiclassical and leaves out effects such as spin-orbit coupling and a full description of the nuclear dipole-dipole interactions (which we approximate as nuclear spin noise), both of which may be important for a complete understanding of the experiments.
%
  
 The main implication of the paper for  DNP experiments in double dots is that the nuclear spin dynamics are dominated by either rapid saturation of polarization or an instability to the growth of large difference fields.  These results are consistent with the experimental  observations reported in Refs.~\onlinecite{Petta08}, \onlinecite{Foletti09} and \onlinecite{Barthel12}; however, we see evidence that the dynamics are much richer as the experiments have not resolved whether or not the instability to large difference fields results from dot asymmetry or coherent electron-nuclear interactions.  These two cases could be experimentally distinguished by measuring the sign of $D_z$ in a given double dot.   Furthermore, we showed that the zero states may be experimentally observable as metastable states in certain parameter regimes, indicating that there is still much to explore in the polarization dynamics of double quantum dots.

\acknowledgments
We  thank S.\ Foletti, H.\ Bluhm, C.\ Barthel, C.\ M.\ Marcus, M.\ Rudner, A.\ Yacoby, and M.\ Stopa for valuable conversations. 
Research was supported  by the Physics Frontier  Center at the Joint Quantum Institute, DARPA, MTO, NSF grant DMR-0908070 and by the Office of the Director of National
Intelligence, Intelligence Advanced Research Projects Activity (IARPA),
through the Army Research Office grant W911NF-12-1-0354.

\appendix

\section{Parameters Used in Simulations}

In table I  below we provide a summary of the parameters used in the simulations for each figure.  While many parameters are chosen to be consistent with experiments, not all those presented are self-consistent or experimentally realistic.  In particular, in Fig.~5e the $\Lambda_0$ parameter is unphysically large and in Figures 4, 5ab and 8 the small $m_0$ values correspond to very large magnetic fields.

\setlength{\tabcolsep}{5pt}
\renewcommand{\arraystretch}{1.45}
\begin{table*}
\caption{Parameters used in the simulations shown in the figures of this paper.}
\begin{center}
\begin{tabular}{|c|c|c|c|c|c|c|c|c|c|c|}
\hline
Fig.& $\Delta_0$&$\Gamma_0$ &$\Delta_-$&$ \Lambda_+$&$\Lambda_0$ &$\Gamma_R$& $m_0$ &$\eta$& $M$  \\
\hline
 3 	& 0.5 		& $0.005-0.5$ 	& $0-0.4$ 		& 0 	& 0 		& 0 					& 0			& 0.005 	& 400\\
\hline
 4 	& $5\cdot m_0$ 	&
 $f_c/2$	& $1.25\cdot m_0$ 	& 5 	& $5\cdot m_0$	& $2.7\cdot\Gamma_0 $		& $10^{-3}-10^{-1}$& 0		& 400\\
\hline
5ab 	& 0.19 	& 1 			& 0.0048 		& 5.8 	& 0.002	& $ 2.7$ 	& $5\cdot 10^{-4}$	& 0 		& 100\\ 
\hline
5c 	& 0.78		& 1			& 0.19			& 5.8	& 0.08		& 1 					& 0.01			& 0		& 100\\ 
\hline
5e 	& 1		& 1			& 0.25			& 0.5	& 1		& 1 					& 0.05			& 0		& 100\\
\hline
6a 	&1.99 	& 1 			& 0.143 		& 626 & 0.5		& $ 2.7$ 	& 0.01			& 0 		& 100\\
\hline
6b 	& 1.99 	& 1 			& 0.143 		& 626 & 0.5		& $ 2.7$ 	& 0.01			& $10^{-4}$	& 100\\
\hline
7a 	& 0.014	& 0.36			& 0.0034		& 5	& 0.014	&  $2.7\cdot \Gamma_0 $	& 0.0027		& $4\cdot 10^{-4}$	& 200\\
\hline
7b 	 &0.013	& 2.1			& 0.0034		& 5	& 0.013	& $2.7\cdot \Gamma_0 $		& 0.0027		& $2\cdot 10^{-3}$	& 200\\
\hline
8 	 &$5\cdot m_0$ 	&
$f_c/2$& $1.25\cdot m_0$ 	& 5 	& $5\cdot m_0$	& $2.7\cdot\Gamma_0 $		& $10^{-3}-10^{-1}$& $10^{-4}$	& 400\\
\hline
9a 	 &0 		& 1			& 0			& 0	& 0		& 0					& 0			& $10^{-3}$	& 200 per species\\
\hline
9b 	 &0.5 		& $0.005-0.5$	&  $0-0.4$ 		& 0	& 0		& 0					& 0			& $5\cdot10^{-5}$	&400 per species\\
\hline
\end{tabular}
\end{center}
\label{tab:GaAsNum}
\end{table*}

\section{$\Phi$ variables}
In this appendix we describe a systematic approach to coarse graining the electron wavefunction in solving the semiclassical equations of motion, which we refer to as the Independent Random Variable Annular Approximation (IRVAA).   We construct a sequence of discretizations of the wavefunction for which we can provide a rigorous bound on the error in time evolution compared to the exact solution.  In the process we also introduce a new set of statistically independent nuclear spin variables, which are a convenient basis for numerical simulations.  

We see from Eqs.~\ref{eq:EOM} and \ref{eq:PsPd} that the semiclassical evolution of each spin depends only on the vectors $\vec{L}$ and $\vec{R}$ (or equivalently on $\Dvec$ and $\Svec$). That is, if we know $\bm{P}_d(t)$ (which depends only on $\vec{L}$ and $\vec{R}$), then we can solve for the dynamics of the entire system. However, even if we know $\bm{P}_d(t)$, if we look at the equation of motion for $\bm{L}$ we find that it generates an infinite hierarchy of equations
\begin{align} \label{eqn:contEOM}
  \frac{d \bm{L}}{dt}=\bm{P}_l \times  \vec{L}^*,
\end{align}
where we defined $\bm{L}^*\equiv\sum_k  g_{kl}^2 \vec{i}_{kl}$. Now $\dot{\bm{L}}^*$ couples to the variable $\sum_k g_{kl}^3 \vec{i}_{kl}$ and so on.


 To find an approximate solution to the dynamics we would like to find an effective method to truncate this infinite hierarchy of equations.  For simplicity we focus on the case where $\bm{P}_\ell$ is only a function of $\bm{L}$, reducing it to a single dot problem, and drop the dot indices in the following discussion.  We also work in the continuum limit, which is defined by a nuclear angular momentum density $\bm{I}(\bm{r},t)= \sum_{k} \vec{i}_k(t) \delta(\bm{r}-\bm{r}_k)$.
 
Each variable in the hierarchy of equations of motion (as in Eq.\ \ref{eqn:contEOM}) can be expressed as an integral
 \be
 \bm{\Phi}(t) = \int d^d r \, g(\bm{r})\, \varphi(g(\bm{r})) \, \bm{I}(\bm{r},t),
 \ee
 where $\varphi(x)$ is a polynomial in $x$. That is, there is a one-to-one correspondence between polynomials $\varphi(x)$ and the variables in the EOM. For example, $\bm{L}$ corresponds to $\phi(x)=1$. 
 
 We would like to think of a truncation procedure as any procedure that provides a reduced, self-consistent set of equations describing the evolution of $\bm{P}$, equivalently $\bm{L}$.  We make a formal definition of a truncation procedure as a procedure producing a set of variables $\bm{\Phi}_k$, $k=1,\ldots,M$, of the form above and an $M\times M$  matrix $Q$, such that $\bm{\Phi}_1 = \bm{L}$ and 
 \beu
 \frac{d \bm{\Phi}_k}{d t} = \sum_{\ell} \bm{P} \times Q_{k\ell} \bm{\Phi}_\ell.
 \eeu
 Since we always constrain $\bm{\Phi}_1=\bm{L}$, we always have $\phi_1(x)=1$.
 
To construct a convenient basis of nuclear spin variables we first define a norm $\mean{\cdot}_{\varphi}$ based on the statistical average of a nuclear spin variable in the infinite temperature ensemble, i.e. 
\begin{align}
\mean{\Phi \cdot \Psi}_{\varphi}& = \int d^d r \, d^d r' \, g^2(r) \varphi(g(r)) \psi(g(r')) \mean{\bm{I}(\bm{r}) \cdot \bm{I}(\bm{r}')}_e  \nonumber \\
&= \frac{I(I+1)}{a^d} \int d^d r\, g^2(r) \varphi(g(r)) \psi(g(r))
\end{align}
 where $a$ is the lattice spacing, $\mean{\cdot}_e$ is the ensemble average over the initial thermal state and we took $\mean{\bm{I}(\bm{r})\cdot \bm{I}(\bm{r}')} = I (I+1) \delta(\bm{r}-\bm{r}')/a^d$.  Now we can construct an orthogonal set of polynomials with respect to this norm by using the standard Gram-Schmidt procedure starting from the polynomial $1$.   This gives a set of orthogonal polynomials $\varphi_k$ and associated nuclear spin variables $\bm{\Phi}_k= \int d^dr\, g(\bm{r}) \varphi_k(g(\bm{r})) \bm{I}(\bm{r},t)$,  which are statistically independent in the infinite temperature ensemble ($\ie, \av{\bm{\Phi}_k\cdot\bm{\Phi}_l}=3\Omega_l^2 \delta_{kl}$)and satisfy $\bm{\Phi}_1=\bm{L}$.  
 
 The equations of motion (EOM) for these variables can be written as
 \be\label{eq:PhidotDef}
\dot{\bm{\Phi}}_n = \bm{P} \times Q_{nm} \bm{\Phi}_m
\ee
where the matrix $Q_{mn}$ is a tridiagonal matrix defined by the recurrence relations
\be
x \varphi_n(x) = Q_{nn-1} \varphi_{n-1}+Q_{nn} \varphi_n + Q_{nn+1} \varphi_{n+1}
\ee
and we used the fact that $x \varphi_n(x)$ only has a non-zero overlap with $\varphi_n$ and $\varphi_{n\pm1}$.
 
 We now define an $M^\text{th}$ order truncation procedure with respect to the variables $\bm{\Phi}_k$ by setting $Q_{M M+1}=0$.   The central result of this appendix is encapsulated by the following theorem for this truncation procedure.

\textbf{Theorem:}~\textit{ For a given wavefunction $g(\bm{r})$ and $\varepsilon>0$, the above truncation procedure at order $M$ will produce an effective $\bm{L}^M(t)$
such that $\abs{\bm{L}(t)-\bm{L^M}(t)}<\varepsilon$ for all $t<t_M$, where $t_M$ is a time scale that increases linearly with $M$ and $\bm{L}(t)$ is the exact result for the untruncated system.}

 We begin our analysis by proving that any truncation procedure is equivalent to a discretization of the function $g(\bm{r})$ (\ie, an annular approximation), by which we mean a representation of $\bm{L}$ as 
 \be  \label{eqn:discL}
 \bm{L} = \sum_{k=1}^M g(\bm{r}_k) \tilde{\bm{I}}_k,
 \ee
 where $\tilde{\bm{I}}_k$ is a rescaled nuclear spin variable associated with position $\bm{r}_k$.  

The reverse implication is clear because if we start with such a discrete representation, then the variable associated with the polynomial 
\beu \label{eqn:wcPoly}
w(x)= \prod_{k=1}^M [x - g(\bm{r}_k)]
\eeu
is identically zero. That is, if there are only $M$ discrete spins in the system, then there are only $M$ statistically independent variables $\bm{\Phi}_k$ in the system, and $\bm{\Phi}_{M+1}$ is naturally zero.  This result naturally truncates Eq.\ \ref{eq:PhidotDef}. Consequently, if we consider any basis of  polynomials of degree less than $M$ and its associated set of spin variables, then we can obtain a finite, self-consistent set of equations for the evolution of $\bm{L}$.  

The forward implication follows along similar lines.  If $M-1$ is the maximal degree of the set of polynomials $\{\varphi_k(x)\}$ associated with the truncation variables $\{\bm{\Phi}_k\}$ and $\Phi_M$ is the spin variable corresponding to this polynomial, then, when we compare to the continuum limit, we find that the statement that $d{\Phi}_M /dt$ does not couple to higher degree polynomial variables implies the existence of a degree-$M$ polynomial $w(x)$ such that
\beu
\int d^d r \,g(\bm{r})\, w(g(\bm{r})) \, \bm{I}(\bm{r},t) = 0,
\eeu
for any $\bm{I}(\bm{r},t)$. The existence of such a polynomial immediately implies that we can represent $\bm{L}$ in the discretized form of Eq.~\ref{eqn:discL}.

We have now reduced the problem of finding an optimal truncation procedure to the problem of finding an optimal discretization procedure for integrals of the form
\beu
\int d^d r \, g(\bm{r}) \varphi(g(\bm{r}))\, \bm{I}(\bm{r},t),
\eeu
where $\varphi(x)$ is a polynomial in $x$.  Fortunately, this last problem is solved through the theory of Gaussian quadrature.  \cite{numAnal}  First, though, we assume that our function $g(\bm{r})$ is spherically symmetric so that we can write our integrals as effective one-dimensional integrals with respect to the rescaled angular momentum density
\be\label{eqn:weight}
\bm{I}(r,t) = \int d \Omega \, a^{d-1}\,N(r)\, \bm{I}(r, \bm{\Omega},t)/S(d)
\ee
where $\bm{\Omega}$ parameterizes the surface of a $d$-dimensional sphere, $a$ is the lattice spacing, $S(d)$ is the surface area of a unit sphere in $d$ dimensions, and $N(r)\equiv S(d)\, r^{d-1}/a^{d-1}$ is the number of nuclear spins at radius $r$; for example, in two dimensions $N(r) = 2 \pi r/a$.  The ensemble average of $\bm{I}(r,t)$ is given by $\mean{\bm{I}(r) \cdot \bm{I}(r')} = I(I+1 )N(r) \delta(r-r')/a $.

To begin constructing our Gaussian quadrature rules we  rewrite 
\be
\begin{split}
\bm{\Phi}(t)&= \int_0^{\infty} dr N(r) g^2(r) \varphi(g(r)) \frac{\bm{I}(r,t)}{N(r) g(r)} \\
&=\int_0^1 dx\, \omega(x) \varphi(x) \frac{\bm{I}\big(g^{-1}(x),t \big)}{N\big(g^{-1}(x)\big) x}
\end{split}
\ee
where $x=g(r)$ and $\omega(x)=\frac{dg}{dr}\lvert_{g^{-1}(x)} N\big(g^{-1}(x)\big) x^2$ is the weight function. 
%
   Standard results in the theory of numerical integration imply the existence of a set of orthogonal polynomials, $\varphi_n$, with respect to the inner product
\be \label{eqn:gaussDot}
(f,h)= \int_0^1 dx\, \omega(x)\, f(x)\, h(x)
\ee
such that, for any function $f(x)$, the $M^\text{th}$ order quadrature approximation is given by 
\be \label{eqn:quad}
\int_0^1 dx \,\omega(x)\, f(x) \approx \sum_{k=1}^M \omega_k \,f(x_k),
\ee
where $x_k$ are the zeros of $\varphi_M$ and the weights  $\omega_k$ are determined by the condition that Eq.~\ref{eqn:quad} is exact for all polynomials of degree strictly less than $2M$.  The error in this formula decreases exponentially in $M$, or better, provided that $f$ is smooth. \cite{numAnal}  In addition, these polynomials are exactly the ones we used to construct our truncation procedure.  
Consequently, our truncation procedure defined above
 is equivalent to 
approximating $\bm{L}$ in quadrature as in Eq.~\ref{eqn:discL} with $\tilde{\bm{I}}_k= w_k \bm{I}(r_k,t)/g_k^2 N(r_k)$.   

To prove the theorem we first note that from the definition  $\abs{\bm{P}(\bm{L})} \le 1$ for all $\bm{L}$.  Now let $p>0$ be such that $\lvert \bm{P}(\bm{L})-\bm{P}(\bm{L}^{'}) \lvert<p  \lvert \bm{L}-\bm{L}^{'} \lvert$ for all $\bm{L}$ and $\bm{L}^{'}$.  We define $\bm{L}_n(t)\equiv\int d^d r g^n(\bm{r}) \bm{I}(\bm{r},t)$ and $\bm{L}_n^M(t)$ is the solution for the equivalent variable in the truncated system of equations.  
To provide bounds on the error propagation we define $\delta^M_n(t)\equiv\lvert \bm{L}_n(t)-\bm{L}_n^M(t) \lvert$. We work in time units where $\max_r g(\bm{r})=1$ and let $b=\max_{n,t} \lvert \bm{L}_n(t) \lvert \le \int d^d r g(\bm{r}) (I+1)$.  Now  it is straightforward to show that
\be \label{eqn:errProp}
\dot{\delta}_n^M \le pb\, \delta^M_1 +(1+p \delta_1^M) \delta_{n+1}^M \le \zeta (\delta^M_1 + \delta^M_{n+1})
\ee
where $\zeta = \max(p b, 1+p \varepsilon)$ and, by assumption, we are restricted to short enough times that $\delta_1^M<\varepsilon$.
By construction, $\delta_n^M(0)=0$ for $n<M$  while for $n>M$ $\delta_n^M$ is bounded by the quadrature error on the integral $\int d^d r g^n(\bm{r}) \bm{I}(\bm{r},0) $, which is less than $c\, e^{-M}$ for a constant $c$ independent of $M$.  Using Eq. \ref{eqn:errProp} we can then bound the error on $\delta_1^M \le c e^{-M}(e^{2 \zeta t}-1)$. This implies that the time to make an error of size  $\varepsilon$ scales as $(1/2\zeta) \log( \varepsilon e^M/c+1) \sim (M-\log c/\varepsilon)/2\zeta$ for large $M$.  This proves the theorem.

For the two dimensional Gaussian $g(r)\propto e^{-r^2/2\sigma^2}$ the weight function $w(x)=x$ and the associated orthogonal polynomials are the Jacobi polynomials.  The matrix $Q$ is then given by standard recurrence relations for Jacobi polynomials.  Once the recurrence relations are known, one can work with the $\Phi$-variables without converting between the original nuclear spin variables because the $\Phi$ variables were defined such that they are initially statistically independent.   This is a convenient numerical approach for these types of central spin problems, and it was used in all of the numerics in this work.  

\section{Multiple Nuclear Species}

In this appendix we include the effects of multiple nuclear species in our simulations and find that the main results for both asymmetric and identical results carry through much the same.  First we show how to include multiple species in terms of the collective $\Phi$-variables and then we present the simulation results.

When multiple species are taken into account we must include the Larmor precession of the nuclear spins.  In this case the EOM take the form
\be
\dot{\bm{I}}_{kd}^\alpha= \gamma_e b_\alpha \, v_0 \abs{\psi_{kd}}^2 \bm{P}_d \times \bm{I}_{kd}^\alpha - \omega_\alpha\, \hat{z} \times \bm{I}_{kd}^\alpha,
\ee
where $\alpha$ is a species index, $\omega_\alpha = \gamma_\alpha \, \Bext T / \tau_a$ is the effective Larmor frequency, $b_\alpha$ is the bare hyperfine field of species $\alpha$, $\gamma_\alpha$ is the gyromagnetic ratio of species $\alpha$, $\Bext$ is the external magnetic field, and we have explicitly included the factor $T/ \tau_a$, where $T$ is the total time of the nuclear pump cycle and $\tau_a$ is  the adiabatic sweep time.

\setlength{\tabcolsep}{5pt}
\renewcommand{\arraystretch}{1.45}
\begin{table}[htdp]
\caption{Relative population of the nuclear species $x_\alpha$, effective hyperfine field due to species $\alpha$ $b_\alpha$, and the gyromagnetic ratio $\gamma_\alpha$, for the three nuclear species in GaAs.}
\begin{center}
\begin{tabular}{|c|c|c|c|}
\hline
&${}^{75}$As&${}^{69}$Ga &${}^{71}$Ga\\
\hline
$x_\alpha$ & 1& 0.6& 0.4 \\
\hline
$b_\alpha$ (T) & -1.84 & -1.52& -1.95 \\
\hline
$\gamma_\alpha$ $\Big(\frac{\textrm{kHz}}{\textrm{mT}}\Big)$& 45.96&64.39 &81.81 \\
\hline
\end{tabular}
\end{center}
\label{tab:GaAsNum}
\end{table}%

\begin{figure}
\includegraphics[width=1.68in]{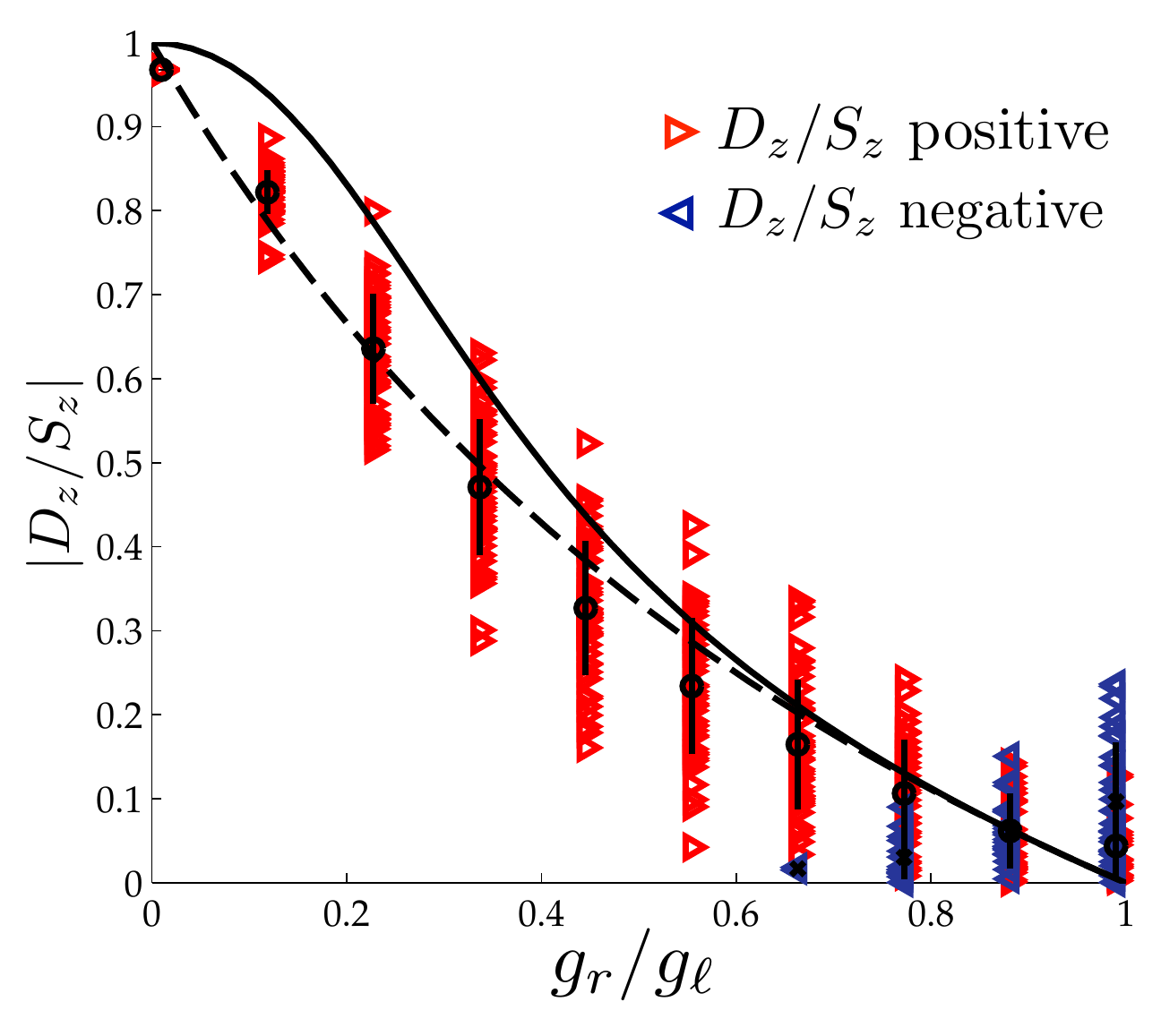}
\includegraphics[width=1.68in]{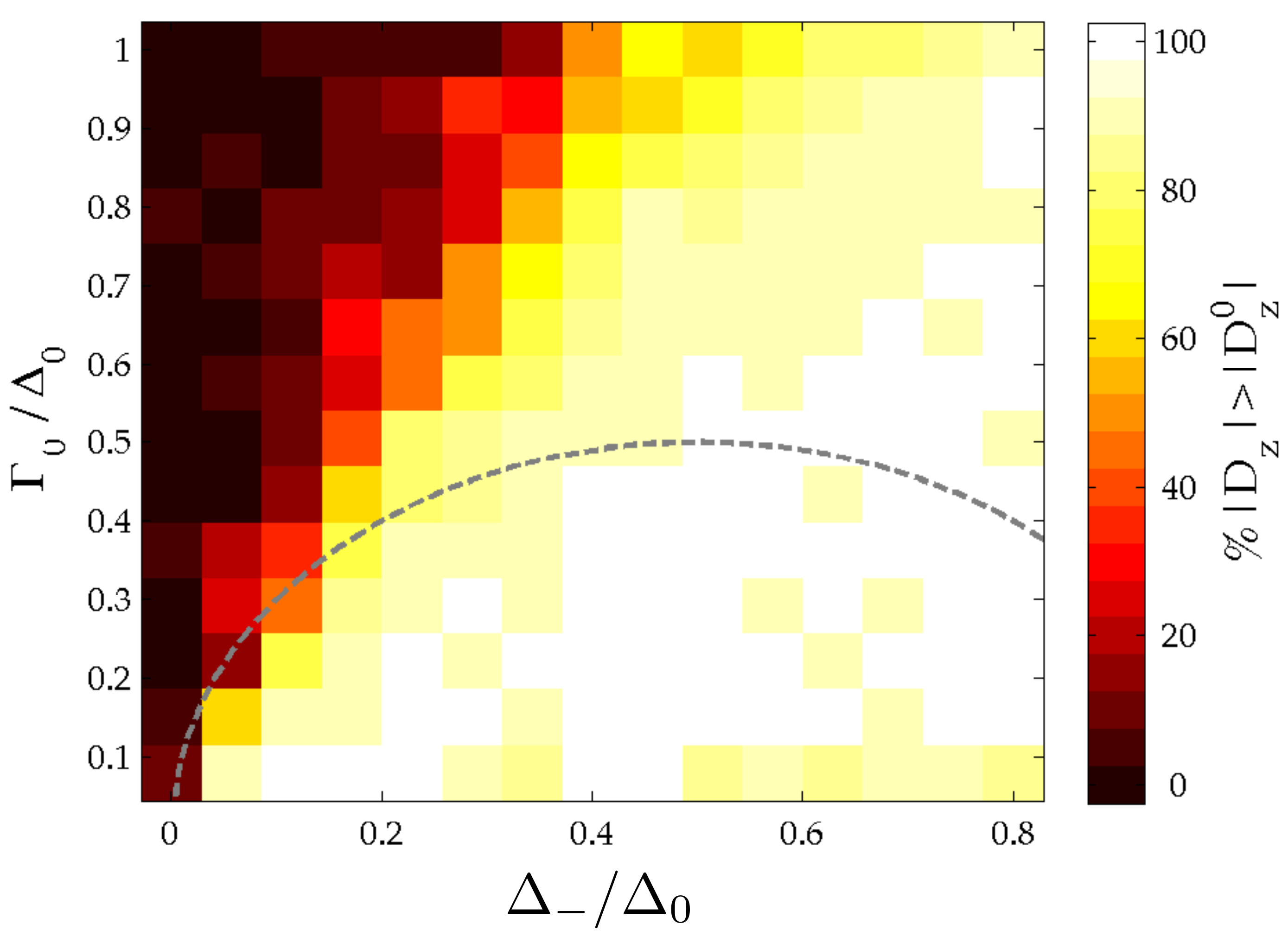}
\caption{\label{fig:mult} a) As in Fig. \ref{fig:ratio}, with parameters chosen as in Fig.~2 of Ref. \onlinecite{Gullans10}, except with three species. Due to the computational cost of running three species of spins, simulations were run for only $10\%$ as long, and the range of $D_z/S_z$ is larger as a result. The trend that $D_z/S_z$ is in good agreement with the single-species prediction is clearly visible. b) Phase diagram with multiple species and $m_0=0$. } 
\end{figure}

We introduce the projector function $\pi_{kd}^\alpha$, such that $\pi_{kd}^\alpha =1$ if there is species $\alpha$ in unit cell $k$ and 0 otherwise.  This allows us to write
\be
\begin{split}
\bm{L} &= \sum_{k,\alpha} \gamma_e b_\alpha v_0 \abs{\psi_{k\ell}}^2
 \pi_{k\ell}^\alpha\, \bm{I}_{k\ell}^\alpha \\
 &= \frac{\Omega_\ell}{\sqrt{\sum_\alpha b_\alpha^2\, x_\alpha}} \sum_{k,\alpha} b_\alpha \,g_{k\ell}\, \pi_{k\ell}^\alpha\, \bm{I}_{k\ell}^\alpha.
 \end{split}
 \ee
  Here we have defined $\Omega_\ell$ to be the standard deviation of $L_\mu$ in the infinite temperature state, explicitly
\begin{align}
\langle {\bm{I}_{kd}^{\alpha}\cdot \bm{I}_{k'd'}^{\alpha'} } \rangle &= I(I+1) \delta_{kk'}\delta_{dd'} \delta_{\alpha \alpha'}, \\
\Omega_\ell^2 \equiv \mean{\bm{L}^2}/3&= \sum_{k,\alpha} \gamma_e^2 b_\alpha^2\,x_\alpha\, v_0^2 \abs{\psi_{k\ell}}^4  \,\frac{I(I+1)}{3} 
\end{align}
where $x_\alpha= \mean{\pi_{kd}^\alpha}$ is the relative proportion of species $\alpha$ on the sites it can occupy, $g_{kd} \propto v_0 \abs{\psi_{kd}}^2$ are chosen to satisfy $\sum_k g_k^2 \,I(I+1) = 3$, and
$I $ is the total spin of a single nuclear spin ($I= 3/2$ for all species in GaAs).  

We define the variables
\be
\bm{\Phi}_{n}^{\alpha} =\frac{1}{\sqrt{x_\alpha}} \sum_k g_{k\ell}\, \varphi_n^\ell(g_{k\ell})\, \pi_{k\ell}^\alpha\, \bm{I}_{k\ell}^\alpha,
\ee
where $\varphi_n^\ell(x)$ are defined as in Appendix B and are independent of the species, i.e.  $\varphi_0^\ell(x)=1$ and 
\be
\sum_k g_{kd}^2\, \varphi_n^\ell(g_{kd})\, \varphi_m^\ell(g_{kd})\, I(I+1) =3\,\delta_{nm}.
\ee
  These definitions have the implication that 
$
\langle {{L}_{n \, \mu}^{\alpha} \cdot{L}_{n'\mu'}^{\alpha' }} \rangle=  \delta_{nn'} \delta_{\mu\mu'} \delta_{\alpha \alpha'},
$ 
and we can draw initial values for each of them from a normal distribution.
Furthermore, we can express 
\be
\bm{L}= \frac{\Omega_\ell}{\sqrt{\sum_\alpha b_\alpha^2\, x_\alpha}} \sum_{\alpha} b_\alpha \sqrt{x_\alpha}\, \bm{\Phi}_{0}^\alpha.
\ee
All these definitions are equivalent for the right dot.

In these variables the EOM take the form
\be
\begin{split}
\dot{\bm{\Phi}}_n^\alpha &=   \frac{ \gamma_e b_\alpha}{N} \, \bm{P}_\ell \times \big( \varepsilon_n \bm{\Phi}_{n-1}^\alpha + \alpha_n \bm{\Phi}_{n}^\alpha \\
&+ \varepsilon_{n+1} \bm{\Phi}_{n+1}^\alpha \big) - \omega_\alpha \, \hat{z} \times \bm{\Phi}_n^\alpha,
\end{split}
\ee
where we have used the definition $N^{-1}= \max_k v_0 \abs{\psi_{kd}}^2$ to represent the number of nuclear spins with which the electron has significant overlap.  For a two dimensional gaussian wave function we have 
$
N= 2/3 \sum_\alpha  x_\alpha  \gamma_e^2 b_\alpha^2 I(I+1)/\Omega^2
$

In Fig.~\ref{fig:mult} we include the three nuclear species in the simulation and show that qualitatively the results from the single species case still hold.  Fig.~\ref{fig:mult}a shows the asymptotic ratio of $D_z/S_z$ as the relative dot sizes are varied, where we see good agreement with the simple prediction given in the introduction.  In Fig.~\ref{fig:mult}b we extract the phase diagram in the simplified model with only $\Delta_{0,-}$ and $\Gamma_0$ non-zero, as in the model of Ref.\ \onlinecite{Gullans10}.   As in the single-spin case, we find a saturation regime at high values of $\Gamma_0/\Delta_0$ and an instability regime at lower values.  Unlike in the single-spin case, the saturation regime does not broaden at higher values of $\Delta_-/\Delta_0$. The dashed line is the same as that in Fig.~\ref{fig:phase1}, showing the simple prediction for the phase boundary with a single species, from Ref.~\onlinecite{Gullans10}.  The lower-left side of the phase diagram (the region most easily reached in experiments) is well-described by this prediction, even with multiple species.

\bibliographystyle{apsrev-nourl}
\bibliography{DNP_DD}

\end{document}